\documentclass{sig-alternate}

\RequirePackage{ifthen}
\usepackage{amsmath,amssymb,url,listings,mathrsfs,wasysym}
\usepackage{tikz}
\usetikzlibrary{shadows}
\usetikzlibrary{shapes}

\usepackage[figtopcap]{subfigure}

\usepackage[all]{xy}
\usepackage{enumerate}
\usepackage{mathtools}

\newenvironment{example}{\textbf{Example 1}}{\qed}
\newenvironment{excont}[1]{\textbf{Example 1 (#1)}}{\qed}

\usepackage{graphicx}
\usepackage{psfrag}

\usepackage{color}

\let\next\undefined
\DeclareMathOperator{\next}{\mathop\bigcirc}

\DeclareMathOperator{\always}{\mathop\square}

\renewcommand{\v}{\varphi}
\newcommand{\e}{\epsilon}
\newcommand{\condition}{\gamma}

\renewcommand{\d}{\delta}
\newcommand{\dropprob}{\Delta}
\newcommand{\D}{\mathcal{D}}
\newcommand{\B}{\mathcal{B}}
\newcommand{\M}{\mathcal{M}}
\newcommand{\R}{\mathcal{R}}
\newcommand{\U}{\mathcal{U}}
\newcommand{\C}{\mathcal{C}}
\newcommand{\E}{\mathcal{E}}
\newcommand{\SE}{\mathcal{E}_S}
\newcommand{\EE}{\mathcal{E}_E}
\newcommand{\G}{\mathcal{E}_G}
\newcommand{\SAP}{\mathfrak{C}}

\newcommand{\Synth}{\mathfrak{S}}
\newcommand{\Realiz}{\mathfrak{R}}
\newcommand{\V}{\mathcal{V}}
\newcommand{\MSG}{\mathrm{MSG}}
\newcommand{\timeout}{\mathrm{T.O.}}
\newcommand{\cond}{\mathrm{cond}}

\newcommand{\locev}[1]{\underline{#1}}

\newcommand{\success}{\locev{\mathrm{success}}}
\newcommand{\fail}{\locev{\mathrm{fail}}}

\newcommand{\vareq}[1]{\overset{#1}{\approx}}

\newcommand{\generated}{\models {\M(\d)}}

\newcommand{\proj}[1]{[\![{#1}]\!]}
\newcommand{\len}[1]{|{#1}|}

\newcommand{\snd}{\mathrm{snd}}

\newcommand{\ack}{\mathrm{ack}}

\newcommand{\nack}{\mathrm{nack}}

\newcommand{\glob}[4]{{#1}_{{#2} \rightarrow {#3}}(#4)}
\newcommand{\eglob}[3]{{#1}_{{#2} \rightarrow {#3}}}
\newcommand{\env}[4]{\locev{#1}_{{#2} \rightarrow {#3}}(#4)}
\newcommand{\eenv}[3]{\locev{#1}_{{#2} \rightarrow {#3}}}
\newcommand{\sys}[4]{\locev{#1}_{{#2} \leftarrow {#3}}(#4)}
\newcommand{\esys}[3]{\locev{#1}_{{#2} \leftarrow {#3}}}

\newcommand{\define}{\sl}

\newcommand{\fig}[1]{Fig.\ \ref{fig:#1}}
\renewcommand{\sec}[1]{Sec.\ \ref{sec:#1}}
\newcommand{\tab}[1]{Table~\ref{tab:#1}}

\newcommand{\figvspace}{\vspace{-0.2in}}
\newcommand{\figsmallvspace}{\vspace{-0.1in}}

\begin{document}

\conferenceinfo{Technical report for the paper with the same title prepared for submission to ICCPS'13,} {April 8--11, 2013, Philadelphia, PA, USA.} 
\CopyrightYear{2013} 
\clubpenalty=10000 
\widowpenalty = 10000

\title{Synthesis of Reactive Protocols for Vehicle-to-Vehicle Communication -- Technical Report\thanks{This work is supported partly by the Studienstiftung des deutschen Volkes, the Boeing Corporation and the AFOSR award number FA9550-12-1-0302.}}%

\numberofauthors{3}

\author{
\alignauthor
Clemens Wiltsche\vspace{0.05in} \\
       \affaddr{University of Oxford, UK}\\[1\jot]
       \email{cw395@cantab.net}
\alignauthor
Ufuk Topcu\vspace{0.05in} \\
       \affaddr{University of Pennsylvania, USA}\\[1\jot]
       \email{utopcu@seas.upenn.edu}
\alignauthor
Richard M.\ Murray\vspace{0.05in} \\
       \affaddr{California Institute of Technology, USA}\\[1\jot]
       \email{murray@cds.caltech.edu} 
}

\maketitle

\begin{abstract}
We present a synthesis method for communication protocols for active safety applications that satisfy certain formal specifications on quality of service requirements. The protocols are developed to provide reliable communication services for automobile active safety applications. The synthesis method transforms a specification into a distributed implementation of senders and receivers that together satisfy the quality of service requirements by transmitting messages over an unreliable medium. We develop a specification language and an execution model for the implementations, and demonstrate the viability of our method by developing a protocol for a traffic scenario in which a car runs a red light at a busy intersection.
\end{abstract}

\category{C.2.2}{Protocol Verification}{}
\category{B.1.2}{Automatic Synthesis}{}

\keywords{Vehicle-to-vehicle communication; Discrete controller synthesis; Active safety}

\vspace{-.06in}
\section{Introduction}

Active safety systems have the potential to transform automobile traffic by complementing a human operator's capabilities to prevent accidents and increase efficiency~\cite{Caveney}. Communication between cars enables cooperative safety applications by further augmenting information gained from local sensors. Examples for active safety applications are traffic signal violation warning, cooperative collision warning and electronic emergency brake light~\cite{Caveney,Sengupta2}.

By transmitting information between each other, cars can gain a view of the traffic situation more refined than it would be possible merely with sensors~\cite{VComm}, because global information about traffic is being made available locally to the cars. Such information can be used in active safety systems to avoid accidents and even improve traffic efficiency by enabling both communication with the infrastructure and between cars~\cite{highways,Caveney,Dresner, collisionwarning, rearend}.

Depending on the active safety application, different types of vehicle-to-vehicle (V2V) communication paradigms have been investigated. Safety applications often require to maintain continuous tracking of other cars in the vicinity, which is typically done by having cars broadcast information about their position, velocity and other parameters of their state in regular intervals~\cite{Sengupta2}. Vehicle Ad-Hoc Networks (VANETs) using routing protocols such as Geocasting or Ad-Hoc Distance Vector (AODV) handle applications involving several cars in a peer-to-peer (P2P) connection~\cite{hartenstein,RAODV,geocasting}. The main challenge is to maintain reliable communication in the presence of possible channel congestion if several cars use the transmission medium simultaneously~\cite{hartenstein, routing}.

Traditionally, in the development of communication protocols, the programs are implemented manually, and verification of the protocol is only done after prototyping, either through testing or model checking~\cite{CSMA, AODVVerify, AdHocVerify}. A slight improvement over this bottom-up approach is to develop a framework for distributed protocol specification and automatically generate inputs to model checkers and theorem provers~\cite{DeclarativeNetwork}. %

In contrast, synthesis finds the programs to be executed on each car directly from a global protocol specification. In this approach, the synthesis method is guaranteed to generate distributed implementations that satisfy their specification by construction. However, so far only small problems have been considered in synthesis without particular applications in mind~\cite{Probert, GlobalClock, Multipath}. Also, protocol implementations are only valuable in practice if it is clear under which assumptions they are correct and if the communication requirements are clearly specified~\cite{hartenstein}. Only if a precise model of the environment is provided, i.e.\ the worst-case behavior of the transmission medium, is an argument of correctness convincing.

Synthesis is made difficult on the one hand by distributing a single global specification in a way that the distributed implementation operates correctly in an adverse environment, and on the other hand by having to ensure correctness of the results, which has to be ensured for \emph{any} valid protocol specification given as input to the synthesis.

Our main contribution is the development of a method that automatically translates global specifications of the protocol into implementations that formally guarantee the desired quality of service requirements under the environment assumptions. Specifically, we develop a synthesis method for reliable asynchronous communication protocols with clearly defined interfaces that can be used in a layered design. We focus on providing communication services to enable active safety applications for cars and therefore lump any active safety activities into an abstract higher level that interacts via strictly defined interfaces with the lower level communication services that we develop. %

Our work addresses several shortfalls in previous work on protocol synthesis~\cite{ChuLiu, Ishida, Probert, Zafiropulo}. We introduce a formal specification language to allow a textual representation of the protocol specifications, which are typically given in graphical form. Moreover, we make precise the semantics of protocol specifications and their implementations that are usually only informally described.

\vspace{-.06in}

\section{Communication} \label{sec:communication}

\begin{figure}
\centering
\psfrag{A}[cc][cc]{\footnotesize{B}}
\psfrag{B}[cc][cc]{\footnotesize{A}}
	\includegraphics[width=0.45\textwidth]{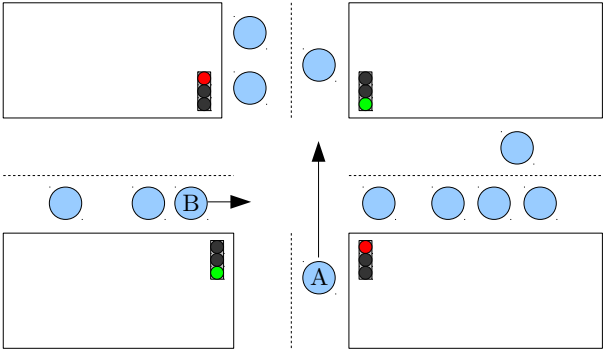}
\caption{Intersection scenario: Car $A$ runs the red traffic light. Hence car $B$ (and potentially other cars) must stop. Other cars at the intersection may be transmitting over the wireless medium at the same time, which has a deteriorating effect on the communication between $A$ and $B$.}
\label{fig:intersection}
\figvspace
\end{figure}

Consider the scenario of two cars wanting to communicate with each other at an intersection, c.f.\ \fig{intersection}. One major complicating factor for reliable V2V communication is that the cars are constantly moving and have to communicate over wireless links. Cars that intend to communicate share restricted bandwidth availability with all other cars within reach. When constantly tracking cars using broadcasts, scalability is limited by the susceptibility for flooding and frequent message collisions~\cite{routing}, and by having to keep track of the state of every car.

It would be desirable to be able to implement communication protocols that guarantee the correct transmission of data even in the presence of a large number of other cars. We therefore consider initiating communication on-demand when required by an active safety application in an emergency. In this approach we do not track every car but only exchange information when required. This approach has the advantage that traffic on the network is lower, more predictable, and reliability guarantees can be provided, as demonstrated in this paper. We adopt an approach in which the sender is responsible for correct delivery by retransmitting data when a package drop is detected~\cite{initiation}.

When considering a V2V communication between two cars, we do not explicitly consider the behavior of all other cars. Since from the point of view of the transceivers it is only relevant whether a message is correctly received, we lump together the behavior of all other cars that are not directly involved in the communication and consider them as a single environment. Our method allows us to explicitly state the assumptions on this environment under which the protocol has to perform correctly.

Communication between cars is governed by a set of rules summarized as a protocol. After a data transfer is initiated, messages are transmitted and received in order to guarantee a reliable delivery. A protocol is implemented by equipping each car with a communication service automaton (CSA), which can be seen as a building block or ``controller'' handling all communication activities. Hence, the protocol can be seen as a building block with clearly defined behavior and interfaces to its environment consisting of higher level active safety components (ASCs) and to the lower level that handles the transmission of the messages over the physical medium.

Each CSA operates locally, i.e.\ it can only interact with the sensors and actuators of the car it is located on. However, since a communication protocol defines events potentially involving several cars, CSAs need to interact with each other. This interaction is done by transmitting messages between the cars e.g.\ using wireless transceivers.

Defining a clear hierarchy of layers is inspired from the ISO OSI architecture prevalent in most modern communication networks~\cite{ISOOSI}: A \emph{network layer} is dedicated to establishing host-to-host connections with basic quality of service (QoS) guarantees. A \emph{data-link layer} is layered below the network layer and provides error-corrected single hop connections. Above the network layer is the \emph{transport layer}, that among other services provides the destination address of a message and QoS requirements. We consider an abstraction in which a car's ASC contains the transport layer and all above layers. The ASC specifies parameters such as the data to be sent, the destination address, and limits on transmission delay.

\vspace{-.06in}

\section{Setup} \label{sec:preliminaries}

Developing a synthesis method requires a formal specification language for protocols and a modelling framework to formally describe executable CSAs. Moreover, the CSAs should include interfaces to their corresponding ASCs at the higher level, and hence our synthesis method is designed to introduce this inter-level interaction. To illustrate our method we will use the following example motivated by Caveney~\cite{Caveney}, and Farkas et al.\ \cite{VComm}:

\begin{example}
Consider the scenario of cars at a road intersection shown in \fig{intersection}. Car $A$ runs a red traffic light, and car $B$ approaches the intersection on a trajectory that would lead to a collision. The two cars have to communicate in order to avoid an accident. At the intersection there might be other cars that share the same broadcast medium and hence might interfere with the communication between $A$ and $B$.
\end{example}

\subsection{Operation of a Protocol} \label{sec:protocol}

\begin{figure}
\centering
\begin{displaymath}
    \xymatrix@R=0.1cm@C=0.5cm{ {\begin{array}{l}\text{\sl Transport}\\\text{\sl Layer and}\\\text{\sl higher}\end{array}} & {\mathrm{ASC}_A} \ar@<1ex>[ddd]^<<<<{\env{\snd}{A}{B}{d}} & & {\mathrm{ASC}_B}  \ar@<1ex>[ddd]^<<<<{\eenv{\ack}{B}{A}} \\ \\ \\ {\begin{array}{l}\text{\sl Network}\\\text{\sl Layer {\phantom{anda}}}\end{array}} & *+++[][F]{M_A} \ar@<1ex>[d]^>>{!!a_{A \rightarrow B}(d)} \ar@<1ex>[uuu]^<<<<{\esys{\ack}{A}{B}} & & *+++[][F]{M_B} \ar@<1ex>[d]^>>{!!b_{B \rightarrow A}} \ar@<1ex>[uuu]^<<<<{\sys{\snd}{B}{A}{d}} \\ {\begin{array}{l}\text{\sl Data-link}\\\text{\sl Layer and}\\\text{\sl below}\end{array}} & {} \ar@<1ex>[u]^<<{?b_{A \leftarrow B}} & {\begin{array}{c}\\\\\text{Medium}\end{array}} \ar@<2.5ex>@{==}[l] \ar@<-2.5ex>@{==}[r] & {}\ar@<1ex>[u]^<<{?a_{B \leftarrow A}(d)}
    }
\end{displaymath}
\figvspace
\caption{Two cars $A$ and $B$ communicating with each other: $A$ sends a message (by calling $\env{\snd}{A}{B}{d}$) and $B$ responds with an acknowledgement on reception (by calling $\eenv{\ack}{B}{A}$). The transmission medium and the ASCs are the environment of the CSAs.}
\label{fig:cars}
\figvspace
\end{figure}
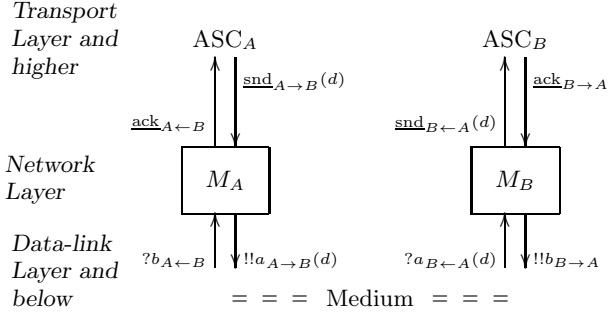

The primary objective of a communication protocol is to transfer information between cars. Information transfer between two cars can be interpreted as synchronizing two local events between the cars. Events are indexed by elements $\e$ from a set $\E$. An event $\e$ may be associated with data $d$ from a set $\D$, written $\e(d)$. $\D$ contains an auxiliary element $\perp$, indicating the absence of data. We simply write $\e$ for notational convencience if $\d = \perp$ in $\e(d)$.

A {\define local event} is an event that is triggered either by the ASC of a car (an {\define environment-triggered event} from the point of view of the CSA), or by the CSA of a car itself (a {\define system-triggered event}). An environment-triggered event that is initiated by the ASC at car $A$ and is to be synchronized with car $B$ is written as $\env{\e}{A}{B}{d}$. It is synchronized with the corresponding system-triggered event $\sys{\e}{B}{A}{d}$ by the CSA of car $B$. The sets of environment-triggered and system-triggered events are written as $\EE$ and $\SE$ respectively.

If $\env{\e}{A}{B}{d}$ on car $A$ is synchronized with $\sys{\e}{B}{A}{d}$ on car $B$, then the data $d$ is transferred from $A$ to $B$. This is summarized as a single {\define global event} $\glob{\e}{A}{B}{d}$ (note the absence of the line under $\e$). The set of global events is denoted by $\G$. A protocol specification defines a desired temporal order on such global events. Since global events involve several cars, a protocol specification is {\define centralized}, i.e., it is assumed that the actions of all cars can be influenced independently by a single controller.

Synchronization is achieved by sending messages across a shared transmission medium. A CSA interacts with the medium by transmitting messages and waiting for reception of messages. A message {\define transmission} is indicated by ``$!!$'', while a {\define reception} is indicated by ``?'' prefixed to a message.

The interaction with the higher-level ASC is managed by {\define calls} and {\define upcalls}.  A call is initiated by the ASC and causes an environment-triggered event in the CSA. An upcall is initiated by a system-triggered event in the CSA.

\begin{excont}{Continued}
Consider again the intersection problem in \fig{intersection}. As car $A$ is approaching the intersection, it needs to establish whether it is safe to enter the intersection. It therefore wants to establish a communication with any car that might pose a safety hazard.

Car $A$ needs to communicate with car $B$ to find out if $B$ is willing and able to stop or whether $A$ should attempt an emergency brake. Each car is assigned a unique address for labelling messages, so that when a car receives a message, it knows whether it is the intended destination. We assume that the ASC at $A$ provides its CSA with the address of $B$, so that a P2P communication with $B$ can be established.

This communication scenario is shown in \fig{cars}, where the CSA associated with each car is shown as a box. Data $d$ is transferred from $A$ to $B$, and $B$ should send an acknowledgement back to $A$. Sending $d$ from $A$ to $B$ is done by synchronizing the local events $\env{\snd}{A}{B}{d}$ and $\sys{\snd}{B}{A}{d}$, while the acknowledging synchronizes $\eenv{\ack}{A}{B}$ with $\esys{\ack}{B}{A}$. A call by an ASC triggers the corresponding environment-triggered event in the CSA on the same car, while an upcall is initiated by the CSA when some system-triggered event requires the attention of the higher level.
\end{excont}

\subsection{Quality of Service} \label{sec:QoS}

Depending on the application, it is be necessary to guarantee that a transmission is completed with certain requirements on particular aspects such as end-to-end delay, message drop probability or bandwidth. These aspects are called Quality of Service (QoS). We are concerned with automatically implementing protocols that guarantee that certain requirements on QoS are met.

Whether QoS requirements can be satisfied depends on the properties of the medium used to transmit messages over. In our work we assume minimal capabilities for a transceiver, so the only way to satisfy QoS requirements is to select the appropriate frequency and number of retransmissions for messages.

Also, when finding the CSAs that satisfy the protocol, we have to take into account that the performance of the transmission medium typically degrades as consequence of messages being transmitted. Moreover, a car cannot predict the behavior of the transmission medium merely on the basis of its own actions, since there might be other cars sharing the same medium that exhibit unpredictable behavior from the point of view of the car. In Example 1, while cars $A$ and $B$ are communicating, other cars might be trying to transmit messages itself, leading to a degradation in performance for $A$ and $B$ that neither car can predict.

We restrict the package drop probability $\dropprob$ of the transmission medium by assuming that it is below a given threshold probability $\d$ at all times. We write this as $\always(\dropprob \leq \d)$, where ``$\always$'' is the always operator ``$\always$'' of linear temporal logic (LTL) Hence, a full specification in the framework can be stated as an assumption/guarantee specification~\cite{AGspec} $\always(\dropprob \leq \d) \rightarrow \v$, where a protocol specification $\v$ only has to hold as long as the assumption that at all times $\dropprob \leq \d$ is satisfied.

A straightforward extension to take time into account would be to consider each (re)transmission to take up some amount of time $\mathcal{T}$. We can then include another assumptions of the form $\always(\mathcal{T} \leq \tau_{max})$, where $\tau_{max}$ is an upper bound on the transmission time.

\vspace{-.06in}

\section{Technical Approach} \label{sec:approach}

In this section the concepts described above are formalized.

\subsection{Protocol Specification Language} \label{sec:speclang}

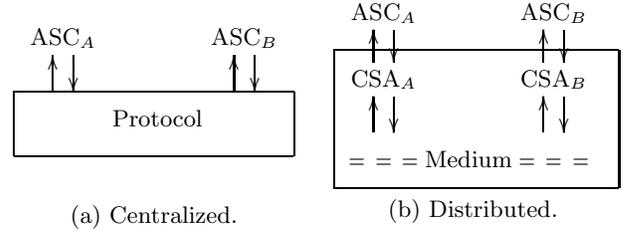
\begin{figure}
\centering
\subfigure[Centralized.]{
    \xymatrix@R=0.0cm@C=-0.1cm{ \\ & {\mathrm{ASC}_A} \ar@<1ex>[ddd] & & {\mathrm{ASC}_B}  \ar@<1ex>[ddd] \\ \\ \\ & {} \ar@<1ex>[uuu] &   & {} \ar@<1ex>[uuu]  \\ & & {\text{ Protocol}} \\ & {} &  & & & {} \save "5,1"."7,5"*[F]\frm{} \restore \\ \\ \\ \\ }}
    \hspace{0.1cm}
\subfigure[Distributed.]{
    \xymatrix@R=0.0cm@C=-0.1cm{& {\mathrm{ASC}_A} \ar@<1ex>[ddd] & & {\mathrm{ASC}_B}  \ar@<1ex>[ddd] \\ \\ \\  & {\mathrm{CSA}_A} \ar@<1ex>[ddd] \ar@<1ex>[uuu] & & {\mathrm{CSA}_B}  \ar@<1ex>[ddd] \ar@<1ex>[uuu] \\ \\ \\ & {} \ar@<1ex>[uuu] &   & {} \ar@<1ex>[uuu] & &  \\ & & {\text{Medium}} \ar@{==}[rr] \ar@{==}[ll] & {} & & \\ & {} &  & & & {} &  \save "3,1"."8,5"*+[F]\frm{} \restore }}
\figsmallvspace
\caption{The protocol on two levels of detail.}
\label{fig:protocol}
\figvspace
\end{figure}

In a protocol specification, the protocol is viewed as a single component interacting with the ASCs, cf.\ \fig{protocol}(a). Only the interaction across the interface between ASCs and CSAs is specified. The CSAs, representing an implementation of a protocol, then interface with the lower level transmission medium in order to provide the required services to the higher level. In this way, the ASCs never come in direct contact with the transmission medium. As introduced above, a specification of the protocol is given as a temporal (partial) order of global events. A global event is any event involving the interaction of several cars, such as a message transmission (involving both the sender and the receiver).

Hence, a protocol specification can be seen as an allowed set of sequences of global events. Moreover, each sequence is tagged with a QoS requirement, which, in our case simply is the required probability of the sequence to be synchronized correctly. In order to avoid having to write a list of sequences with potentially many global events repeating, we use the following temporal logic-like language to define {\define protocol specifications}:
\begin{equation*}
	\v ::=  e^p | e \rightarrow \next\v | \v \vee \v,
\end{equation*}
where $e^p$ is a global event $e \in \G$ together with a probability $p$ indicating the required QoS. Extensions of this specification language can also include time, bandwidth or other QoS requirements in the same way in the specification. A {\define specification} is a protocol specification together with an environment assumption, in our case an upper bound $\d$ on the drop probability $\dropprob$.

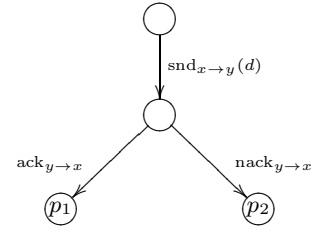
\begin{figure}
\centering
\begin{displaymath}
\xymatrix{ & *++[o][F-]{} \ar[d]^{\snd_{x \rightarrow y}(d)} \\ & *++[o][F-]{} \ar[dl]_(0.7){\ack_{y \rightarrow x}} \ar[dr]^(0.7){\nack_{y \rightarrow x}} \\ *+[o][F-]{p_1} & & *+[o][F-]{p_2}}
\end{displaymath}
\figvspace
\caption{Visualization of protocol specification $\v$ in \eqref{eq:protspec}, which establishes a partial order between the global events $\snd_{x \rightarrow y}(d)$, $\ack_{y \rightarrow x}$ and $\nack_{y \rightarrow x}$, represented by the order of the edge labels.}
\label{fig:protspec}
\figvspace
\end{figure}

\begin{excont}{Continued}
The protocol described in the intersection example of \fig{intersection} can be specified as
\begin{equation}
	\v = \glob{\snd}{A}{B}{d} \rightarrow \next (\eglob{\ack}{B}{A}^{p_1} \vee \eglob{\nack}{B}{A}^{p_2}), \label{eq:protspec}
\end{equation}
which can be illustrated as a tree as in \fig{protspec}. In the numerical results presented later for this example we will use different values for $p_1$ and $p_2$. 

The results of the synthesis also depend on the drop probability bound $\d$. A complete specification that includes the assumptions on the transmission medium dynamics would be
\begin{equation}
	\always(\dropprob \leq \d) \rightarrow \v.
\end{equation}
\end{excont}

Since we are interested in QoS requirements over the drop probability of the transmission medium, a probability $p$ on $e^p$ labels each leaf of the tree representing a protocol specification, specifying the desired probability of the (unique) sequence of global events $\sigma$ occurring that leads to the leaf. We call a sequence $\sigma$ with a probability $p$ attached to it a {\define $p$-sequence} and write $(\sigma)^p$. The semantics of the protocol specification language is defined by a satisfaction relation: If a $p$-sequence $\sigma$ of global events satisfies the protocol specification $\v$, this is written as $(\sigma)^p \models \v$.

We first develop an intuitive understanding of a sequence $\sigma = e_1e_2\ldots$ satisfying a specification $\v$. Recall that a protocol specification only takes the interfaces between CSAs and ASCs into account, and hence views the protocol implementation as a monolithic entity as in \fig{protocol}(a). Each global event $e_i = \glob{\e}{x}{y}{d}$ in the sequence $\sigma$ is interpreted as the synchronization of an environment-triggered event $\env{\e}{x}{y}{d}$ and a system-triggered event $\sys{\e}{y}{x}{d}$. The ASC on car $x$ triggers $\env{\e}{x}{y}{d}$ by a call to its CSA. The intention is that the corresponding system-triggered event $\sys{\e}{y}{x}{d}$ is synchronized with that event in the CSA on car $y$ (and an upcall is made to its ASC). The synchronization is correct if after an environment-triggered event $\env{\e}{x}{y}{d}$, the first system-triggered event is $\sys{\e}{y}{x}{d}$, i.e.\ no other system-triggered event is interleaved between them. Note that environment-triggered events that do not correspond to global events in the specification may be interleaved, as the protocol has no control over the higher level. Then, the statement $(\sigma)^p \models \v$ expresses that $\sigma$ satisfies the partial order defined in $\v$ and has a high enough probability $p$ attached to it.

We now formally define $\models$ recursively on the structure of a protocol specification $\v$ (cf.\ \eqref{eq:protspec}):
\begin{equation}
\begin{array}{rlll}
	(e)^p &\models e^q &\Leftrightarrow & p \geq q \\
	(e, \sigma)^p &\models e \rightarrow\next\v &\Leftrightarrow & (\sigma)^p \models \v \\
	(\sigma)^p &\models \v_1 \vee \v_2 &\Leftrightarrow &(\sigma)^p \models \v_1 \mathrm{~or~} (\sigma)^p \models \v_2,
\end{array}
\end{equation}
where adding a global event $e$ to the head of a sequence $\sigma$ is written as $e, \sigma$. Under these semantics a protocol specification is satisfied exactly by those sequences of global events that both obey the partial order induced by $\varphi$ and that have a sufficiently high probability attached to them.

\subsection{Communication Service Automata} \label{sec:CSA}

\begin{figure}[hbt!]
\begin{center}
\subfigure[Sender: $M_A$]{
        \xymatrix@C=0.8cm@R=1cm{ & *++[o][F=]{s_1} \ar[d]^{\env{\snd}{A}{B}{d}}="snd" \\
     *++[o][F-]{s_4} & *++[o][F-]{s_2} \ar[l]_(.5){\scriptsize\begin{array}{l}\fail_1 \\ \nu_1 > n_1\end{array}} \ar@/^/[d]^(.7){!!a_{A \rightarrow B}(d)}="!!a"^(.5){\nu_1 \leq n_1} \\
       & *++[o][F-]{s_3} \ar[dl]_(.7){?b_{A \leftarrow B}}="?b"_(.5){\esys{\ack}{A}{B}}="rack" \ar[dr]^(.7){?c_{A \leftarrow B}}="?c"^(.5){\esys{\nack}{A}{B}}="rnack" \ar@/^/[u]^(.3){\timeout_1}^(.5){\nu_1+\!+} \\
       *++[o][F.]{s_5} & & *++[o][F.]{s_6} }}
\subfigure[Receiver: $M_B$]{
    \xymatrix@C=0.8cm@R=1cm{ {} & &  *++[o][F=]{s_1} \ar[d]_{?a_{B \leftarrow A}(d)}="?a"^{\sys{\snd}{B}{A}{d}}="rcv" & & {}\\
     & & *++[o][F-]{s_2} \ar[dl]_(.5){\eenv{\ack}{B}{A}}="ack" \ar[dr]^(.5){\eenv{\nack}{B}{A}}="nack" & & {} \\
      *++[o][F-]{s_4} & *++[o][F-]{s_3} \ar@/^/[d]^(0.7){!!b_{B \rightarrow A}}="!!b"^(.5){\nu_2 \leq n_2} \ar[l]_(.5){\scriptsize\begin{array}{l}\fail_2 \\ \nu_2 > n_2\end{array}}="fail2" & & *++[o][F-]{s_7} \ar@/_/[d]_(0.7){!!c_{B \rightarrow A}}="!!b"_(.5){\nu_3 \leq n_3} \ar[r]^(.5){\scriptsize\begin{array}{l}\fail_3 \\ \nu_3 > n_3\end{array}}="fail3" & *++[o][F-]{s_8} \\
       & *++[o][F]{s_5} \ar@/^/[u]^(.3){?a_{B \leftarrow A}(d)}^(.5){\nu_2+\!+} \ar[d]_(.65){\timeout_2}="?b"_(.5){\success_2}="success2" & & *++[o][F]{s_9} \ar@/_/[u]_(.3){?a_{B \leftarrow A}(d)}_(.5){\nu_3+\!+} \ar[d]^(.65){\timeout_3}="?b"^(.5){\success_3}="success3" \\
        & *++[o][F.]{s_6} & & *++[o][F.]{s_{10}}}}
\end{center}
\figvspace
\caption{Two CSAs that realize the protocol specification in~\eqref{eq:protspec}. The transitions are labelled with broadcast messages (e.g.\ $!!a_{A \rightarrow B}(d)$), receptions (e.g.\ $?a_{B \leftarrow A}(d)$), and local events (e.g.\ $\sys{\snd}{A}{B}{d}$ and $\env{\snd}{B}{A}{d}$). Initial states and final states are shown as doubled and dotted circles respectively. In each retransmission loop, a retransmission counter $\nu_i$ is increased by one on each timeout or reception. The transmissions are conditioned on the retransmission counters $\nu_i$. If the counter is exceeded, indicated by $\nu_i > n_i$, a $\fail_i$ occurs.}
\label{fig:services}
\figvspace
\end{figure}
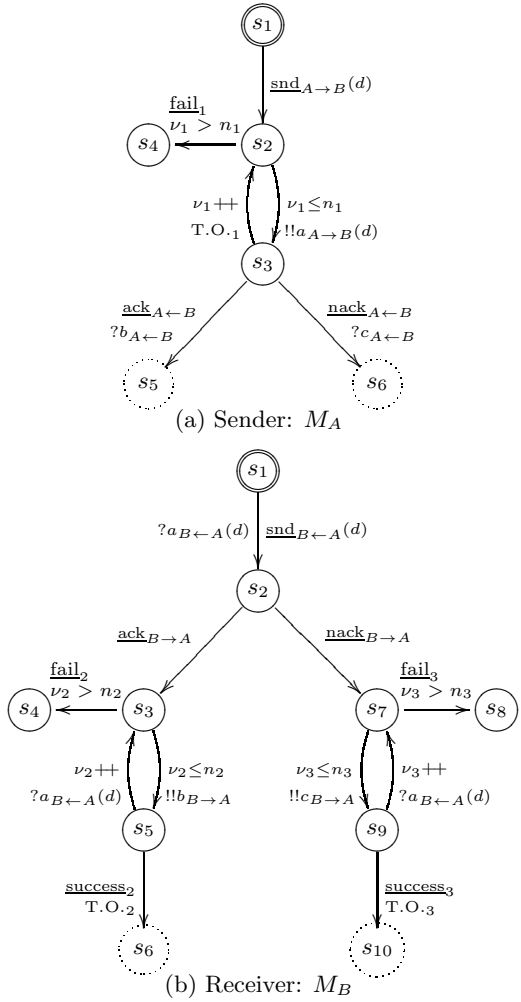

We are interested in finding a way to implement a protocol in a distributed manner by finding CSAs for the cars so that their joint execution satisfies the protocol specification. That is, the implementation of the protocol must use the transmission medium in order to guarantee the required services to the higher level, see \fig{protocol}(b). A set $\M$ of CSAs satisfies a protocol specification if it produces only the allowed sequences of global events, and these with a sufficiently high probability. In this section we make precise the concept of a CSA and define its semantics in the next section.

A CSA is a finite state machine with labelled transitions, which is similar to a protocol entity specification used by Ishida et al.\ \cite{Ishida}. Transition labels either indicate which actions should be executed when the transition is taken, or impose conditions on a transition. A transition for which all conditions are satisfied is called {\define enabled}. The labels that are available to the synthesis method are explained below, and transitions are typically labelled with combinations of labels.

Firstly, interaction with the higher-level ASCs is encoded by edges labelled with environment-triggered and system-triggered events. We also introduce two special system-triggered events ``$\fail$'' and ``$\success$'' to $\SE$ that have no corresponding environment-triggered events to be synchronized with. The purpose of these events is merely to inform the ASC of the outcome of a transmission. A $\fail$ event indicates that allowable retransmission count is exceeded, while the $\success$ event indicates a successful transmission to the ASC triggering the last global event.\footnote{This is necessary since no response from another ASC can indicate completion of the transmission.}

Secondly, to interact with the transmission medium, transitions can be labelled with message transmissions and receptions. Each message has a unique identifier $m$. A broadcast message is written as $!!m_{x \rightarrow y}(d)$, where $x$ and $y$ are the source and destination respectively, and $d \in \D$ is the data transmitted. It is read as ``send $m$ with data $d$ to $y$ from source $x$.'' Similarly, a reception is written as $?m_{y \leftarrow x}(d)$, where $x$, $y$ and $d$ have the same interpretation as for a broadcast message. It is read as ``receive $m$ with data $d$ from $x$ destined for $y$.'' Again, if $d = \perp$, the parameter is not written. Define $\B$ and $\R$ to be the set of broadcasts and receptions respectively.

Lastly, we introduce labels for internal actions of a CSA. In order to satisfy the QoS requirements of the protocol specification, it may be necessary to allow the retransmission of messages. To this end, we define a set of {\define variables} $\V$ over $\mathbb{N}_0 = \{0, 1, 2, \ldots \}$ that act as {\define retransmission counters}. We will construct the CSAs in such a way that for each message $m$ that might be retransmitted, after a transmission $!!m_{x \rightarrow y}(d)$, either a reception of some other message is expected or a {\define timeout} ``$\timeout$'' may occur. On the timeout, the retransmission counter $\nu$ of the message is increased by one. If $\nu$ exceeds its {\define retransmission bound} $n \in \mathbb{N}_0$, the transmission fails, causing a $\fail$ event and a corresponding upcall informing the ASC. We write an {\define update} of a variable $\nu \in \V$ as $\nu\!+\!+$, and denote the set of updates by $\U = \{ \nu\!+\!+ | \nu \in \V\}$. Further, a transition may be labelled by a {\define condition} on a retransmission counter, which can be either of the form $\nu \leq n$ or $\nu > n$. The set of conditions is defined as $\C = \{\nu \bowtie n | \nu \in \V, \bowtie \in \{ \leq, > \}, n \in \mathbb{N}_0 \}$

When synthesizing a CSA from a protocol specification, each transition can be of one of seven kinds, depending on the labels: A environment-triggered event, a conditional system-triggered event, a timeout with a system-triggered event, a timeout with update, a conditional broadcast, a reception with a system-triggered event or a reception with update. Hence, the set of labels is $\Sigma = \EE \cup (\SE \times \C) \cup (\{\timeout\} \times \SE) \cup (\{\timeout\} \times \U) \cup (\B \times \C) \cup (\R \times \SE)  \cup (\R \times \U)$ for the respective cases. The set of such transition labels is denoted by $\Sigma$. A CSA $M$ is a quintuple
\begin{equation*}
	M \triangleq \langle S, \V, s^{init}, S^f, T \rangle,
\end{equation*}
where $S$ is a set of {\define states} labelled by valuations of variables $\V$, $s^{init} \in S$ is the {\define initial state}, $S^f \subseteq S$ is the set of {\define final states}, and $T : S \times \Sigma \rightarrow S$ is the (partial) {\define transition function}.

\begin{excont}{Continued}
The pair of CSAs shown in \fig{services} represents one potential implementation of the protocol specification in \eqref{eq:protspec}. The transmissions and receptions are introduced in order to ensure that the QoS requirements as defined in the specification is preserved by the CSAs that can only communicate over the transmission medium. For example, the sender $M_A$ may retransmit the message $!!a_{A \rightarrow B}(d)$ up to $n_1$ times in case of repeated timeouts to increase the likelihood of a successful transmission, in order to meet the specification.
\end{excont}

\subsection{Semantics of CSAs}

In the semantics of a CSA, we want to reflect that a car should be able to execute it as a controller for its wireless transceiver.

Decisions when to make transitions should be based only on information available locally. For example, a transition labelled by a reception $?m_{x \leftarrow y}$ is taken only when a message $m$ arrives that has $x$ as its destination and $y$ as its source. Since a CSA is executed locally on a car, we first define the {\define local semantics} of a single CSA. This describes how a CSA operates in isolation when receiving calls from the ASC on the same car, and messages from the transmission medium, cf.\ \fig{protocol}(b). We then define the {\define global semantics} of several CSAs that operate together, which requires to take the transmission medium dynamics into account, cf.\ \sec{QoS}. Hence, the global semantics can be interpreted as defining the behavior of the protocol in \fig{protocol}(a). An example of how the semantics are used is presented in \sec{Pdeduction}.

\subsubsection{Deduction Rules}

For ease of presentation, the semantics are defined as a set of {\define deduction rules}. A deduction rule is of the form
\begin{equation*}
\frac{
\begin{array}{cccc}
	H_1 & H_2 & \ldots & H_n
\end{array}}{
C},
\end{equation*}
which is the same as $\bigwedge_{i = 1}^n{H_i} \Rightarrow C$, i.e.\ the {\define conclusion} $C$ follows from the {\define hypotheses} $H_1, H2, \ldots, H_n$. A deduction rule can be applied if all its hypotheses hold. Rules can either be applied forward, starting from one or several axioms, or backwards, starting from a conclusion. Forward application corresponds to simulation, while backwards application corresponds to verification.

\subsubsection{Notation}

\newcommand{\smallstep}[2]{\xrightarrow{\hspace{0.3cm}{#1}\hspace{0.3cm}}_{M_{#2}}}
\newcommand{\bigstep}[1]{\xRightarrow{\hspace{0.3cm}{#1}\hspace{0.3cm}}_{\M(\d)}}

We first introduce some notation to make the statement of the rules more compact. Retransmission uses conditional transitions and updating of variables. The value $v$ of a variable $\nu \in \V$ in a state $s$ is written $s(\nu) = v$. In the initial state $s^{init}$ all variables valuate to zero. A condition $\condition = \nu \bowtie n$ is satisfied in state $s$, written $\condition(s)$, if and only if $s(\nu) \bowtie n$. Two states $s$ and $s'$ are equivalent on their values of the variables in $V \subseteq \V$, written $s \vareq{V} s'$, if and only if $\forall \nu \in V . s(\nu) = s'(\nu)$. We write $T(s, \varsigma) \vareq{V} s'$ if and only if $T(s, \varsigma) = s'$ and $s \vareq{V} s'$, where $\varsigma \in \Sigma$ may stand for any transition label. Furthermore, we use the $+$ operator to append an element to the end of a sequence.

\subsubsection{Local Semantic Rules}

\newcommand{\lsp}{3.7ex}

\begin{table}
\centering
\begin{equation*}
\hspace{-0.3cm}\begin{array}{ll}

\text{[env]} &
\frac{
\begin{array}{c}
T(s, \env{\e}{y}{z}{d}) \vareq{\V} s'
\end{array}}{
\begin{array}{c}
\langle \rho, s \rangle \smallstep{\mathrm{e}}{y} \langle \rho + \env{\e}{y}{z}{d}, s' \rangle
\end{array}} \\[\lsp]

\text{[sys-c]} &
\frac{
\begin{array}{cc}
T(s, (\sys{\e}{y}{z}{d}, \condition)) \vareq{\V} s' & \condition(s)
\end{array}}{
\begin{array}{c}
\langle \rho, s \rangle \smallstep{\mathrm{e}}{y} \langle \rho + \sys{\e}{y}{z}{d}, s' \rangle
\end{array}} \\[\lsp]

\text{[to-sys]} &
\frac{
\begin{array}{c}
T(s, (\timeout, \sys{\e}{y}{z}{d})) \vareq{\V} s'
\end{array}}{
\begin{array}{c}
\langle \rho, s \rangle \smallstep{\mathrm{t}}{y} \langle \rho + \timeout + \sys{\e}{y}{z}{d}, s' \rangle
\end{array}} \\[\lsp]

\text{[to-upd]} &
\frac{
\begin{array}{cc}
T(s, (\timeout, \nu\!+\!+)) \vareq{\V \backslash \{\nu\}} s' & s'(\nu) = s(\nu) + 1
\end{array}}{
\begin{array}{c}
\langle \rho, s \rangle \smallstep{\mathrm{t}}{y} \langle \rho + \timeout, s' \rangle
\end{array}} \\[\lsp]

\text{[b-c]} &
\frac{
\begin{array}{cc}
T(s, (!!m_{y \rightarrow z}(d), \condition)) \vareq{\V} s' & \condition(s)
\end{array}}{
\begin{array}{c}
\langle \rho, s \rangle \smallstep{\mathrm{e}}{y} \langle \rho + !!m_{y \rightarrow z}(d), s' \rangle
\end{array}} \\[\lsp]

\text{[r-sys]} &
\frac{
\begin{array}{cc}
T(s, (\varsigma, \sys{\e}{y}{z}{d})) \vareq{\V} s' & \varsigma = ?m_{y \leftarrow z}(d)
\end{array}}{
\begin{array}{c}
\langle \rho + \varsigma, s \rangle \smallstep{\mathrm{r}}{y} \langle \rho + \varsigma + \sys{\e}{y}{z}{d}, s' \rangle
\end{array}} \\[\lsp]

\text{[r-upd]} &
\frac{
\begin{array}{cc}
T(s, (?m_{y \leftarrow z}(d), \nu\!+\!+)) \vareq{\V \backslash \{\nu\}} s' & s'(\nu) = s(\nu) + 1
\end{array}}{
\begin{array}{c}
\langle \rho + ?m_{y \leftarrow z}(d), s \rangle \smallstep{\mathrm{r}}{y} \langle \rho + ?m_{y \leftarrow z}(d), s' \rangle
\end{array}}

\end{array}
\end{equation*}
\caption{Local semantic rules for deducing behavior of a single CSA $M_y$.}
\label{tab:singleCSArules}
\figvspace
\end{table}

The local semantics is defined by a relation $\smallstep{}{} \subseteq (\Sigma^* \times S) \times (\Sigma^* \times S)$ between sequences of transition labels and CSA states. The statement $\langle \rho, s \rangle \smallstep{}{} \langle \rho', s' \rangle$ means that $M$ at state $s$ transforms $\rho$ into $\rho'$ by making a single transition to state $s'$. It holds if and only if it is deducible via the rules given in \tab{singleCSArules}. To make the statement of the global semantics simpler, we may label the relation by a superscript to distinguish which rules are applied. For example, $\smallstep{\mathrm{e}}{}$ indicates that either the rule [env], [sys-c] or [b-c] are applied. If the superscript is omitted, any rule may be applied.

We explain the [env] rule for CSA $M_y$ in detail, the other rules are similar. The hypothesis $T(s, \env{\e}{y}{z}{d}) \vareq{\V} s'$ expresses that $M_y$ must allow a transition from $s$ that is labelled with the environment-triggered event $\env{\e}{y}{z}{d}$ and leads to a state $s'$ in which the values of all variables in $\V$ are the same as in $s$ (i.e.\ there is no update). If this hypothesis is satisfied, $M_y$ at state $s$ transforms $\rho$ into $\rho'$ by making a transition to state $s'$.

The [env] rule can be applied at any point if a transition labelled by an environment-triggered event is enabled. It is not dependent on an input from the higher level ASC. Stating the rule this way is sufficient for our presentation, but it can be substituted by
\begin{equation*}
\begin{array}{ll}
\text{[env$'$]} &
\frac{
\begin{array}{c}
T(s, \env{\e}{y}{z}{d}) \vareq{\V} s'
\end{array}}{
\begin{array}{c}
\langle \rho + \env{\e}{y}{z}{d}, s \rangle \smallstep{\mathrm{e}}{y} \langle \rho + \env{\e}{y}{z}{d}, s' \rangle
\end{array}}
\end{array}
\end{equation*}
to explicitly require an input to be able to apply the rule. The input is the last element in the sequence, which is an environment-triggered event $\env{\e}{y}{z}{d}$, indicating that the ASC must have made the corresponding call. This input may be placed in the sequence (i.e.\ added as last element) by the global semantics, similar to the inputs for the [r-sys] and [r-upd] rules.

The [sys-c] rule places no restriction on the input and contains as an additional hypothesis that the condition $\gamma$ must be satisfied in state $s$. The system-triggered event $\sys{\e}{y}{z}{d}$ gives rise to an upcall. Such outputs can be read off the last element of the deduced sequence and hence are not modelled explicitly in these rules. The [to-sys] rule can be applied for a transition labelled with a timeout $\timeout$ and a system-triggered event $\sys{\e}{y}{z}{d}$. In the [to-upd] rule the value of the variable $\nu \in \V$ is incremented by one as the transition is taken. Hence we use the operator $\vareq{\V \backslash \{\nu\}}$, since $\vareq{\V}$ would indicate that all variables in $\V$ retain their values as the transition is taken. The [b-c] rule can be applied for a conditional broadcast message. The outgoing message again can be obtained from the last element of the deduced sequence. Rules [r-sys] and [r-upd] require the reception $?m_{y \leftarrow z}(d)$ to occur, hence the rules require the corresponding input.

Each CSA may deduce a set of sequences of events by transitioning between its states. Decisions between environment-triggered events, receptions and timeouts are made by inputs (or the absence thereof) received either from the higher level ASC or the lower level transmission medium. These inputs can only be generated by the global semantics.

\subsubsection{Transmission Medium Modelling}

We define the global semantics by modelling how the transmission medium operates. That is, we define the behavior of the protocol in \fig{protocol}(a) by composing the behavior of the CSAs in \fig{protocol}(b) and abstracting away all lower level detail. The global semantics defines when inputs to a CSA are received from the transmission medium, and restricts the valid interleavings of locally generated sequences. The medium therefore also acts as an arbiter or scheduler of transitions.

In the global semantics, we are interested in ensuring that several CSAs together satisfy the global protocol specification by interacting with each other. We therefore define the semantics of a list of CSAs $\M = \langle M_A,  M_B \ldots \rangle$ that is executed together on the respective set of cars $\SAP = \{A, B, \ldots \}$. Each execution starts with all CSAs in $\M$ being in their initial state $s^{init} = \langle s_A^{init}, s_B^{init}, \ldots \rangle$ and making only transitions allowed by the semantics. Only a single sequence $\rho \in \Sigma^*$ is deduced, which is an interleaving of the sequences deduced locally. The deduction rules also express that the medium transmits messages only with a given probability. Hence, the deduced sequence $\rho$ is tagged with a probability $p$, indicating how likely it occurs.

Not only do the global semantics define how messages are transmitted, also the valid interleavings of locally deduced sequences are restricted. To motivate this, consider in \fig{services} the execution of the environment-triggered event $\eenv{\ack}{B}{A}$. Since time is abstracted away, the transition may be delayed by an arbitrary amount of time. However, then the retransmission loop in the sender $A$ cannot reliably increase the likelihood of a successful execution, since the timeout transition can also be taken at any time. In order to prevent this from happening, the global semantics ensure that transitions that are not timeouts or receptions are taken immediately if enabled. Hence, only one CSA is allowed to make transitions until a timeout or reception is encountered. Then any CSA may make a transition. This is incorporated in the global semantics by always prioritizing one CSA is to make a transition. If this CSA has no transition enabled, any other CSA may make a transition.

\subsubsection{Global Semantic Rules}

\begin{table}[bt!]
\centering
\begin{equation*}
\hspace{-0.5cm}\begin{array}{ll}

\text{[trans]} &
\frac{
\begin{array}{c}
\langle \rho + ?m_{z \leftarrow y}(d), s_z \rangle \smallstep{\mathrm{r}}{z} \langle \rho', s_z' \rangle \\ \varsigma = !!m_{y \rightarrow z}(d)
\end{array}}{
\begin{array}{c}
\langle (\rho + \varsigma)^p, s, y \rangle \bigstep{} \langle (\rho')^{(1-\d)p}, s[z \leftarrow s_z'], z \rangle
\end{array}} \\[\lsp]

\text{[drop]} &
\frac{
\begin{array}{c}
\langle \rho + ?m_{z \leftarrow y}(d), s_z\rangle \smallstep{\mathrm{r}}{z} \langle \rho', s_z' \rangle \\ \varsigma = !!m_{y \rightarrow z}(d)
\end{array}}{
\begin{array}{c}
\langle(\rho + \varsigma)^p, s, y \rangle \bigstep{} \langle (\rho)^{\d p}, s, z \rangle
\end{array}} \\[\lsp]

\text{[nacc]} &
\frac{
\begin{array}{c}
\neg(\langle \rho + ?m_{z \leftarrow y}(d), s_z \rangle \smallstep{\mathrm{r}}{z} \langle \rho', s_z' \rangle) \\ \varsigma = !!m_{y \rightarrow z}(d) \hspace{1cm} \langle \rho, s_x \rangle \smallstep{}{x} \langle \rho', s_x' \rangle
\end{array}}{
\begin{array}{c}
\langle (\rho + \varsigma)^p, s, y \rangle \bigstep{} \langle (\rho)^p, s, z \rangle
\end{array}} \\[\lsp]

\text{[pr-e]} &
\frac{
\begin{array}{ccc}
\varsigma \neq !!m_{y \rightarrow z}(d) & \langle \rho + \varsigma, s_y \rangle \smallstep{\mathrm{e}}{y} \langle \rho', s_y' \rangle
\end{array}}{
\begin{array}{c}
\langle (\rho + \varsigma)^p, s, y \rangle \bigstep{} \langle (\rho')^p, s[y \leftarrow s_y'], y \rangle
\end{array}} \\[\lsp]

\text{[pr-t]} &
\frac{
\begin{array}{cl}
\varsigma \neq !!m_{y \rightarrow z}(d) & \neg(\langle \rho + \varsigma, s_y \rangle \smallstep{\mathrm{e}}{y} \langle \rho'', s_y'' \rangle) \\ & \langle \rho + \varsigma, s_y \rangle \smallstep{\mathrm{t}}{y} \langle \rho', s_y' \rangle
\end{array}}{
\begin{array}{c}
\langle (\rho + \varsigma)^p, s, y \rangle \bigstep{} \langle (\rho')^p, s[y \leftarrow s_y'], y \rangle
\end{array}} \\[\lsp]

\text{[npr]} &
\frac{
\begin{array}{cl}
\varsigma \neq !!m_{y \rightarrow z}(d) & \neg(\langle \rho + \varsigma, s_y \rangle \smallstep{\mathrm{e, t}}{y} \langle \rho'', s_y' \rangle) \\ & \langle \rho + \varsigma, s_x \rangle \smallstep{}{x} \langle \rho', s_x' \rangle
\end{array}}{
\begin{array}{c}
\langle (\rho + \varsigma)^p, s, y \rangle \bigstep{} \langle (\rho')^p, s[x \leftarrow s_x'], x \rangle
\end{array}}
		
\end{array}
\end{equation*}
\caption{Global semantics for deducing behavior of several CSAs $\M$. In particular, the rules [trans], [drop] and [nacc] model the transmission medium.}
\label{tab:compositionrules}
\figvspace
\end{table}

In the global semantics, we are interested in ensuring that several CSAs together satisfy the global protocol specification by interacting with each other. The transmission medium therefore acts as an arbiter or scheduler of transitions. Hence, we can think of the global behavior of several CSAs $M_1, M_2, \ldots$ as an interleaving $\rho$ of the locally generated sequences $\rho_1, \rho_2, \ldots$ of the respective CSAs. Messages are only transmitted with a certain probability. Hence, the sequence $\rho$ is tagged with a probability $p$, indicating how likely it occurs.

The relation $\langle (\rho)^p, s, x \rangle \bigstep{} \langle (\rho')^{p'}, s', x' \rangle$  defines the global semantics according to the rules in \tab{compositionrules}. It means that $\M$ with drop probability $\d$ at state $s$ transforms $\rho$ into $\rho'$ by making a transition to state $s'$ while the priority changes from $M_x$ to $M_{x'}$. In the statement of the rules, updating the $z^{\mathrm{th}}$ element $s_z$ in state $s = \langle s_A, s_B, \ldots, s_z, \ldots \rangle$ with $s_z'$ is written as $s[z \leftarrow s_z']$.

The [trans], [drop] and [nacc] rules define the transmission medium dynamics. If the last deduced element in the sequence is a broadcast message, i.e.\ $\varsigma = !!m_{y \rightarrow z}(d)$, the medium tries to transmit. An application of the [trans] rule models a successful message transmission. This only occurs if the CSA for which the message was destined, $M_z$ makes a transition labelled with the corresponding reception. That is, $\langle \rho + ?m_{z \leftarrow y}(d), s_z \rangle \smallstep{\mathrm{r}}{z} \langle \rho', s' \rangle$ is only satisfied if $M_z$ can execute [r-sys] or [r-upd]. Since a message transmission occurs with probability $1-\d$, the probability with which the sequence $\rho'$ is tagged in the conclusion of [trans] is $(1-\d)p$.

An application of the [drop] rule models a dropped message. It has exactly the same hypotheses as [trans], but its conclusion reflects that no progress has been made. The sequence $\rho$ is tagged with $\d p$ due to the message drop probability $\d$. Note that the priority is at the source CSA $M_y$, which may now execute a timeout transition (if enabled).

The [nacc] rule is applied when a message should be transmitted, but the destination CSA has no transition enabled that is labelled by the corresponding reception. Similar to the [drop] rule, no progress is made. Also, the probability of the deduced sequence is not affected.

The [pr-e], [pr-t] and [npr] rules may be applied if the last element of the sequence is not a message transmission. Then the transmission medium is inactive, and and the CSA that is currently prioritized may execute: If a transition that is not a timeout or reception is enabled, then [pr-e] is applied. If a timeout transition is enabled, then [pr-t] is applied. The [npr] rule may only applied if the currently prioritized CSA has no such transitions enabled. In this case, any CSA $M_x$ may execute.

The transitive closure $\langle (\rho)^p, s, x \rangle \bigstep{}^* \langle (\rho')^{p'}, s', x' \rangle$ denotes that $\langle \rho, s \rangle$ is transformed into $\langle \rho', s' \rangle$ in an arbitrary number of deduction steps. The CSAs $\M$ execute by starting in state $s^{init}$ with an empty 1-sequence $(\bullet)^1$ and any CSA $M_x$ prioritized. Valid deductions are the tuples $\langle (\rho)^{p}, s, y \rangle$ for which $\langle (\bullet)^1, s^{init}, x \rangle \bigstep{}^* \langle (\rho)^{p}, s, y \rangle$.

Note that an example of how the local and global semantic rules are used is included in \sec{Pdeduction}.

\subsubsection{Global Event Sequences}

By applying the deduction rules, sequences over both envi\-ronment-triggered and system-triggered events, broadcasts, receptions and timeouts can be obtained from a set of CSAs. Since protocol specifications are over global events, we need to extract the synchronizations of local events in the sequences generated by a set of CSAs. We therefore define the projection function $\proj{\cdot} : \Sigma^* \rightarrow \G^*$ to find a sequence over global events from $\rho$. It is defined by
\begin{align*}
\proj{\bullet} &\triangleq \bullet\\
\proj{\rho + r} &\triangleq \begin{cases}
	\proj{\rho} + \env{\e}{y}{z}{d} &\text{if}~r = \env{\e}{y}{z}{d} \\
	\proj{\rho'} + \glob{\e}{y}{z}{d}&\text{if}~\proj{\rho} = \proj{\rho'} + \env{\e}{y}{z}{d}\\
	&\hspace{0.5cm}\text{and}~r = \sys{\e}{z}{y}{d}\\
	\proj{\rho} &\text{otherwise}.
\end{cases}
\end{align*}
We use the projection function $\proj{\cdot}$ to express whether a set of CSAs satisfies a protocol specification $\v$ under the environment assumptions $\always(\dropprob \leq \d)$.

\subsection{Correctness} \label{sec:correctness}

In this section we define correctness of a protocol's implementation in form of a set of CSAs $\M$ with respect to a specification $\always(\dropprob \leq \d) \rightarrow \v$. If $\M$ satisfies this specification this is written as $\M \vdash \always(\dropprob \leq \d) \rightarrow  \v$. Correctness depends on the probability of sequences $\sigma$ being synchronized correctly by the CSAs $\M$ if the transmission medium's drop probability $\dropprob$ is bounded from above by $\d$, i.e.\ it satisfies $\always(\dropprob \leq \d)$. If this assumption on the transmission medium is not satisfied, the specification $\always(\dropprob \leq \d) \rightarrow \v$ is trivially satisfied by any set of CSAs. However, this case is useless in practice, as the protocol will not deliver the data with the required QoS.

\subsubsection{Definitions}

Given a protocol specification $\v$, correctness of an implementation depends on whether all CSAs involved in synchronizing a sequence of global events are in a final state. We therefore define the set of {\define globally final} states $S^f_\M$ to include all tuples of states $\langle s_A, s_B, \ldots \rangle \in \prod_{x \in \SAP}S_x$ so that if there is some sequence involving CSAs $x, y, \ldots$, the states $s_x, s_y, \ldots$ are actually final states from $S_x^f, S_y^f, \ldots$.%

We say that a $p$-sequence $(\rho)^p$ is {\define generated} by a set of CSAs $\M$ and drop probability $\d$, and write $(\rho)^p \generated$, if it can be deduced by the rules in \tab{singleCSArules} and \tab{compositionrules} and the deduction ends in a globally final state $s^f \in S^f_{\M}$. Formally,
\begin{align*}
(\rho)^p \generated \Leftrightarrow \\
\exists s^f \in S^f_{\M} . \exists x, y \in \SAP . \langle (\bullet)^1, s^{init}, x \rangle \bigstep{}^* \langle (\rho)^p, s^f, y \rangle.
\end{align*}
As noted above in \sec{protocol}, a $p$-sequence $(\sigma)^p$ satisfies a specification $\v$ exactly if the probability $p$ that all (global) events in $\sigma$ are correctly synchronized is high enough given that the corresponding environment-triggered events are all triggered through calls by the higher level ASCs. For a set of CSAs therefore to satisfy a specification, it is required that the synchronization of events in each sequence is performed with high enough probability.

\subsubsection{Correctness Condition}

The important criterion for correctness is not whether a sequence $\rho$ is generated, but whether the QoS requirements are satisfied. This is because the decisions between environment-triggered events (which essentially generate the sequence) are made by the higher level ASCs, over which a CSA has no control. For example, in \fig{services}, the receiver CSA has no control over whether $\eenv{\ack}{B}{A}$ or $\eenv{\nack}{B}{A}$ is triggered by its ASC in state $s_2$. In our case the only QoS requirement is the probability of all global events being correctly synchronized, so the question for correctness becomes: Given that the ASCs trigger the events necessary to generate $\sigma$, how likely is it that all synchronizations performed?

$\M$ might generate a given sequence $\sigma$ in many different ways, since several sequences $\rho$ deducible by the rules in \tab{singleCSArules} might satisfy $\proj{\rho} = \sigma$. For a sequence $\sigma$, we evaluate the sum of all probabilities $p$ for distinct sequences $\rho$ that satisfy
\begin{equation*}
	\cond(\sigma, p, \d, \M) \triangleq (\proj{\rho} = \sigma \wedge (\rho)^p \generated),
\end{equation*}
and get the probability
\begin{equation*}
	r(\sigma, \d, \M) \triangleq \sum_{\cond(\sigma, p, \d, \M)}{p},
\end{equation*}
expressing the likelihood of the events in the sequence $\sigma$ being correctly synchronized when executing all CSAs in $\M$ in parallel (i.e. using the global semantics). {\define Correctness} then is expressed by
\begin{equation*}
	\M \vdash \always(\dropprob \leq \d) \rightarrow \v  \Leftrightarrow  \forall \sigma . (\exists q . (\sigma)^q \models \v) \Rightarrow (\sigma)^{r(\sigma, \d, \M)} \models \v,
\end{equation*}
i.e.\ if $\sigma$ is a sequence allowed by the specification $\v$, $\M$ synchronizes the events $\sigma$ at least as likely as it is required.\\
The algorithmically challenging part in establishing correctness is to evaluate $r(\sigma, \d, \M)$. However, we only need to compute this for the CSAs that we are synthesizing.

\vspace{-.06in}

\section{Synthesis} \label{sec:method}

The protocol synthesis method $\Synth$ translates a specification into a set of CSAs that is guaranteed to satisfy the specification. The inputs to the synthesis are a protocol specification $\v$, a set of cars $\SAP$ and the specification on the transmission medium dynamics $\always(\dropprob \leq \d)$. $\Synth(\v, \SAP, \d)$ produces a CSA for each car $x \in \SAP$ that interacts with the higher level ASCs as outlined in \sec{preliminaries}.

\subsection{Realizability and Well-Posedness}

Synthesis is preceded by a realizability check, i.e.\ checking whether a specification \emph{can} be implemented. That is, checking realizability amounts to deciding whether there exists a set of CSAs that satisfies the specification $\always(\dropprob \leq \d) \rightarrow \varphi$. If a protocol specification $\v$ is realizable for a set of cars $\SAP$ under a drop probability $\d$, this is written as $\Realiz(\v, \SAP, \d)$.

Checking realizability consists of two parts: Firstly, the specification itself must be well-posed, i.e.\ $\v$ must admit a ``reasonable'' implementation in the form of CSAs. Secondly, it must be possible to find retransmission bounds so that the QoS requirements are satisfied under the given drop probability $\d$.

Well-posedness is a purely syntactic requirement on the specification. We introduce this concept because it is easy to check and simplifies the presentation of the synthesis algorithm. A protocol specification $\v$ is {\define well posed} if on every $p$-sequence satisfying $\v$, two ASCs take turns in triggering the events, and there are at least two events on each path through the tree induced by the specification.

These rather strict requirements on the specifications for well-posedness can be relaxed by generalizing the synthesis method presented in the next section appropriately. For example, a straightforward relaxation would be to allow protocol specifications in which for any disjunction $\v_1 \vee \v_2$, the system-triggered events corresponding to the immediately following global events are all triggered by the same ASC.

We do not develop a separate test for realizability but rather show how our method fails for well-posed but nonrealizable specifications.

\subsection{Synthesis Algorithm}

The synthesis method is implemented in two parts: First, the retransmission bounds are calculated. Then the CSAs are constructed using the retransmission bounds. The retransmission bounds are calculated with the structure of the resulting CSAs in mind, so we present the CSA construction first.

For any specification $\always(\dropprob \leq \d) \rightarrow \v$ and any set of cars $\SAP$, if the specification is realizable, the resulting set $\M$ of CSAs from the synthesis, $\Synth(\v, \SAP, \d)$ must satisfy $\always(\dropprob \leq \d) \rightarrow \varphi$. Formally,
\begin{equation*}
	\forall \v . \forall \SAP . \forall \d \in [0, 1] . \Realiz(\v, \SAP, \d) \Rightarrow (\Synth(\v, \SAP, \d) \vdash \always(\dropprob \leq \d) \rightarrow \v).
\end{equation*}
The synthesis method $\Synth$ is implemented in two parts: First, the retransmission bounds are calculated. Then the CSAs are constructed using the retransmission bounds. The retransmission bounds are calculated with the structure of the resulting CSAs in mind, so we present the CSA construction first.

\subsubsection{CSA Construction}

\tab{synthesis} shows the algorithm {\sc{Synthesize}}$(\v, x, i, E, n_{\v})$. This algorithm constructs the CSA $M_x$ for car $x$ from the specification $\v$. The parameter $i$ is used to uniquely index states in the CSA, and $E$ is a set of global events that is used to construct appropriate criteria for retransmission (explained below). $n_{\v}$ is the list of retransmission bounds calculated in the first step (cf.\ \sec{retransmission}).

Each global event $e = \glob{\e}{x}{y}{d}$ that occurs in the protocol specification $\v$ is assigned an environment-triggered event $\env{\e}{x}{y}{d}$, a system-triggered event $\sys{\e}{y}{x}{d}$, a message $m_{\e} \in \MSG$, a variable (as retransmission counter) $\nu_{\e} \in \V$, a retransmission bound $n_{\e}$ from $n_{\v}$, and system-triggered events $\timeout_{\e}$ and $\fail_{\e}$.

The algorithm is invoked by {\sc{Synthesize}}$(\v, x, 0, \emptyset, n_{\v})$, for each car $x \in \SAP$:\footnote{Note that $x$ does not need to occur in the protocol specification $\v$.} It synthesises a CSA for the well-posed protocol specification $\v$ for car $x$, where states are indexed starting from $0$, no previous events are stored ($E = \emptyset$) and the retransmission bounds $n_{\v}$ are used.

{\sc{Synthesize}} recursively decomposes $\v$ into its subparts. If $\v = \v_1 \vee \v_2$, then two CSAs $M_1$ and $M_2$ are constructed from $\v_1$ and $\v_2$ first and joined together by forming the union of their state spaces, final states and transitions and substituting the initial state $s^{init}_{M_2}$ by the initial state $s^{init}_{M_1}$. For this purpose we define $M[s_1/s_2]$ to be the CSA $M$ with all occurrences of $s_2$ substituted by $s_1$.

If $\v = \glob{\e}{y}{z}{d} \rightarrow \next \v'$, the set $E$ of global events that has last been received on the path through the CSA is updated first. Then again the CSA $M$ for $\v'$ is constructed. Depending on which car $x$ the CSA is constructed for, different transitions are now introduced. If $x = y$ then the ASC on car $x$ is responsible for triggering the event $\env{\e}{y}{z}{d}$, and a retransmission loop is introduced:

\begin{minipage}{0.1\textwidth}
(I)
\end{minipage}
\begin{minipage}{0.2\textwidth}
\begin{equation*}
\xymatrix@C=1.2cm@R=1.3cm{ *++[o][F=]{s_i} \ar[r]^{\env{\e}{y}{z}{d}}
     & *+[o][F-]{s_{i+1}} \ar[r]_{\fail_{\e}}^{\nu_{\e} > n_{\e}} \ar@/_/[d]_(.6){!!a_{\e, y \rightarrow z}(d)}_(.45){\nu_{\e} \leq n_{\e}} & *+[o][F-]{s^{init}_{M}}\\
       & *+[o][F-]{s_{i+2}} \ar@/_/[u]_(.4){\timeout_{\e}}_(.55){\nu_{\e}+\!+} & }
\end{equation*}
\end{minipage}

\noindent If $x = z$, then car $x$ synchronizes $\e$ by the system-triggered event $\sys{\e}{z}{y}{d}$:

\begin{minipage}{0.1\textwidth}
(II)
\end{minipage}
\begin{minipage}{0.2\textwidth}
\begin{equation*}
\xymatrix@C=1.5cm@R=1.3cm{*++[o][F-]{s_i} \ar[r]^{\sys{\e}{z}{y}{d}} & *+[o][F-]{s^{init}_M}}
\label{eq:2}
\end{equation*}
\end{minipage}

\noindent In any other case, simply the CSA for $\v'$ is returned as then the car $x$ is not directly involved in the transmission.

Finally, if $\v = \e_{y \rightarrow z}^p(d)$ then no recursive call to {\sc{Synthesize}} is necessary, but a CSA is directly constructed. If $x = y$ then a retransmission loop is constructed:

\begin{minipage}{0.1\textwidth}
(III)
\end{minipage}
\begin{minipage}{0.2\textwidth}
\begin{equation*}
\xymatrix@C=1.2cm@R=1.3cm{ *++[o][F-]{s_i} \ar[r]^{\env{\e}{y}{z}{d}}
     & *+[o][F-]{s_{i+1}} \ar[r]_{\fail_{\e}}^{\nu_{\e} > n_{\e}} \ar@/_/[d]_(.6){!!m_{\e, y \rightarrow z}(d)}_(.45){\nu_{\e} \leq n_{\e}} & *+[o][F-]{s_{i+3}}\\
       & *+[o][F-]{s_{i+2}} \ar@/_/[u]_(.4){?\mu_{y \leftarrow z}(d)}_(.55){\nu_{\e}+\!+} \ar[r]^{\timeout_3}="?b"_{\success_3}="success3" & *+[o][F.]{s_{i+4}}}
\label{eq:3}
\end{equation*}$\Synth$
\end{minipage}

\noindent In this case, a retransmission is not triggered by a timeout, because $\e_{y \rightarrow z}(d)$ is the last global event in a sequence of required synchronizations and no feedback from the car $z$ can be expected. Therefore, a retransmission is initiated by receiving the last message $\mu$ from car $z$ again, because this indicates that $z$ has not received the message $m_{\e}$ correctly. The message $\mu$ is taken from $E$, the set of global events that has last been received on the path through the CSA. Only if no such message is received is a timeout transition made, which indicates success by an upcall to the ASC. The global semantics of CSAs were carefully constructed so that this timeout is only taken if no message $\mu$ is received.

If $x = z$, then car $x$ synchronizes $\e$ by the system-triggered event $\sys{\e}{z}{y}{d}$:

\begin{minipage}{0.1\textwidth}
(IV)
\end{minipage}
\begin{minipage}{0.2\textwidth}
\begin{equation*}
\xymatrix@C=1.5cm@R=1.3cm{*++[o][F-]{s_i} \ar[r]^{\sys{\e}{z}{y}{d}} & *+[o][F.]{s_{i+1}}}
\label{eq:4}
\end{equation*}
\end{minipage}

\noindent In any other case a trivial CSA with one state is returned.

\begin{excont}{Continued}
The resulting CSAs from synthesizing the specification in \eqref{eq:protspec} are shown in \fig{services}. The CSAs were generated by calling {\sc{Synthesize}}$(\v, x, 0, \emptyset, n_{\v})$ for $x \in \{A, B\}$. The retransmission bounds $n_{\v}$ are calculated as explained in the next section according to the QoS requirements and to the bound on the drop probability $\d$.
\end{excont}

\newcommand{\tb}{\hspace{0.7cm}}

\begin{table}[ht]
\begin{tabular}{l}
\sc{Synthesize}$(\v, x, i, E, n_{\v})$\\
\tb\textbf{If} $\v = \v_1 \vee \v_2$ \textbf{Then}\\
\tb\tb$(M_1, i_1) =$~\sc{Synthesize}$(\v_1, x, i, E, n_{\v})$\\
\tb\tb$(M_2, i_2) =$~\sc{Synthesize}$(\v_2, x, i_1, E, n_{\v})$\\
\tb\tb\textbf{Return} $(\langle S_{M_1} \cup S_{M_2}, s^{init}_{M_1}, S^f_{M_1} \cup S^f_{M_2},$\\
\tb\tb\tb\tb\tb\tb $T_{M_1} \cup T_{M_2} \rangle [s^{init}_{M_1} / s^{init}_{M_2}] , i_2)$\\
\tb\textbf{Else If} $\v = \glob{\e}{y}{z}{d} \rightarrow \next \v'$ \textbf{Then}\\
\tb\tb\textbf{If} $\exists \mu, d . \mu_{y \rightarrow z}(d) \in E$ \textbf{Then}\\
\tb\tb\tb\textbf{Replace} $\mu_{y \rightarrow z}(d)$ \textbf{By} $m_{\e, y \rightarrow z}(d)$ \textbf{In} $E$\\
\tb\tb\textbf{Else}\\
\tb\tb\tb\textbf{Insert} $m_{\e, y \rightarrow z}(d)$ \textbf{Into} $E$\\
\tb\tb\textbf{If} $x = y$ \textbf{Then}\\
\tb\tb\tb$(M, i') =$~\sc{Synthesize}$(\v', x, i+3, E, n_{\v})$\\
\tb\tb\tb\hspace{-1cm}(I)$\left\{\begin{array}{l}
T_{M}(s_i, \env{\e}{y}{z}{d}) := s_{i+1}\\
T_{M}(s_{i+1}, (!!m_{\e, y \rightarrow z}(d), \nu_{\e} \leq n_{\e})) := s^{init}_{M}\\
T_{M}(s_{i+1}, (\fail_{\e}, \nu_{\e} > n_{\e})) := s_{i+2}\\
T_{M}(s^{init}_{M}, (\timeout_{\e}, \nu_{\e}\!+\!+)) := s_{i+1}
\end{array}\right.$\\
\tb\tb\tb\textbf{Return} $(\langle S_{M} \cup \{s_i, s_{i+1}, s_{i+2} \}, s_i, S^f_{M}, T_{M} \rangle, i')$\\
\tb\tb\textbf{Else If} $x = z$ \textbf{Then}\\
\tb\tb\tb$(M, i') =$~\sc{Synthesize}$(\v', x, i+1, E, n_{\v})$\\
\tb\tb\tb\hspace{-1cm}(II)$\left\{\hspace{0.1cm}\begin{array}{l}
T_{M}(s_i, (?m_{\e, z \leftarrow y}(d), \sys{\e}{z}{y}{d})) := s^{init}_{M}
\end{array}\right.$\\
\tb\tb\tb\textbf{Return} $(\langle S_{M} \cup \{s_i \}, s_i, S^f_{M}, T_{M} \rangle, i')$\\
\tb\tb\textbf{Else}\\
\tb\tb\tb\textbf{Return} \sc{Synthesize}$(\v', x, i, E, n_{\v})$\\
\tb\textbf{Else If} $\v = \v = \e_{y \rightarrow z}^p(d)$ \textbf{Then}\\
\tb\tb\textbf{If} $x = y$ \textbf{Then}\\
\tb\tb\tb$\mu_{y \leftarrow z}(d) \in E$\\
\tb\tb\tb\hspace{-1cm}(III)$\left\{\begin{array}{l}
T(s_i, \env{\e}{y}{z}{d}) := s_{i+1}\\
T(s_{i+1}, (!!m_{\e, y \rightarrow z}(d), \nu_{\e} \leq n_{\e})) := s_{i+2}\\
T(s_{i+1}, (\fail_{\e}, \nu_{\e} > n_{\e})) := s_{i+3}\\
T(s_{i+2}, (?\mu_{y \leftarrow z}(d), \nu_{\e}\!+\!+)) := s_{i+1}\\
T(s_{i+2}, (\timeout_{\e}, \success_{\e})) := s_{i+4}
\end{array}\right.$\\
\tb\tb\tb\textbf{Return} $(\langle \{s_i, s_{i+1}, s_{i+2}, s_{i+3}, s_{i+4} \}, s_i,$\\
\tb\tb\tb\tb\tb\tb $\{ s_{i+4} \}, T \rangle, i+5)$\\
\tb\tb\textbf{Else If} $x = z$ \textbf{Then}\\
\tb\tb\tb\hspace{-1cm}(IV)$\left\{\hspace{0.1cm}\begin{array}{l}
T(s_i, (?m_{\e, z \leftarrow y}(d), \sys{\e}{z}{y}{d})) := s_{i+1}
\end{array}\right.$\\
\tb\tb\tb\textbf{Return} $(\langle\{s_i, s_{i+1} \}, s_i, \{ s_{i+1} \}, T \rangle, i+2)$\\
\tb\tb\textbf{Else}\\
\tb\tb\tb\textbf{Return} $(\langle \{s_i \}, s_i, \{ s_i \}, \emptyset \rangle, i+1)$\\
\end{tabular}
\caption{Pseudocode of synthesis algorithm. The CSA is constructed from the diagrams explained in the text and referred to by Roman numerals.}
\label{tab:synthesis}
\figvspace
\end{table}

\subsubsection{Retransmission Bounds} \label{sec:retransmission}

Each global event $\e_{x \rightarrow y}(d)$ gets assigned a unique message $m_{\e} \in \MSG$ and a unique retransmission bound $n_{\e} \in \mathbb{N}_0$. The retransmission bounds are evaluated according to the QoS requirements defined in the protocol specification $\v$.

Recall that the protocol specification $\v$ induces a tree, cf.\ \fig{protspec}. Each edge of this tree is translated by the synthesis into a retransmission loop in the CSA of exactly one car, with a retransmission bound associated with that loop. The retransmission bounds have to be selected so that correctness as defined in \sec{correctness} is guaranteed.

We use the semantics to find the conditions on the retransmission bounds that are sufficient for correctness. We can exploit the tree-like structure of the synthesized CSAs: Apart from the last two retransmission loops in each sequence, the message associated with a retransmission loop is never used at a later point in the same sequence.

Each sequence of global events $\sigma = \e_1\e_2\ldots\e_l$ is associated with a sequence of retransmission bounds $n_{\sigma} = n_{\e_1}n_{\e_2}\ldots n_{\e_l}$. Depending on the values of the retransmission bounds, the sequence $\sigma$ is generated correctly with a certain {\define synchronization probability} $P^{\sigma}(n_{\sigma})$ that depends on all retransmission bounds associated with any event in $\sigma$. Note that this probability is the likelihood of $\sigma$ being generated correctly \emph{given} that the calls are made by the ASCs that generate $\sigma$.

\subsubsection{Deduction of $P^{\sigma}(n_{\sigma})$} \label{sec:Pdeduction}

The synchronization probability $P^{\sigma}(n_{\sigma})$ is evaluated as follows: The case of only one retransmission bound ($\len{n_\sigma} = 1$) never occurs. The case of exactly two retransmission bounds ($\len{n_\sigma} = 2$) means that the last retransmission loop uses the message transmitted in the retransmission loop one before last, cf.\ \fig{services}. The example in this figure can be used to deduce the general expression for $P^{\sigma}(n_1, n_2)$. This is because the synthesis will always generate the same pattern for the last two global events in a specification.

We use the sequence $\sigma = \snd_{A \rightarrow B}(d)\ack_{B \rightarrow A}$ to deduce $P^{\sigma}(n_1, n_2)$ by evaluating the probability of correct synchronization by applying the deduction rules in \tab{singleCSArules} and \tab{compositionrules}. The CSAs $\langle A, B \rangle = \M$ start in the initial state $\langle s^A_1, s^B_1 \rangle$. We omit writing the values of the retransmission counters within the states in our presentation. We let $P^{\sigma}(n_1, n_2) = p^{\sigma}_{1, 1}(n_1, n_2)$, where $p^{\sigma}_{i, j}(n_1, n_2)$ are an auxiliary functions describing the probability of reaching a globally final state from $\langle s^A_i, s^B_j \rangle$ when the calls in $\sigma$ are made.

Initially only $M_A$ can execute by applying the [env] rule locally. Hence, we start the deduction at $\langle (\bullet)^1, \langle s^A_1, s^B_1 \rangle, A \rangle$. We find $p^{\sigma}_{1, 1}(n_1, n_2)$ by applying the global rules to deduce all sequences for which $\M$ ends in a globally final state and the calls in $\sigma$ are made. First, apply [pr-e] globally and [env] locally and deduce
\begin{equation*}
	\langle (\bullet)^1, \langle s^A_1, s^B_1 \rangle, A \rangle \bigstep{} \langle (\env{\snd}{A}{B}{d})^1, \langle s^A_2, s^B_1 \rangle, A \rangle
\end{equation*}
globally from
\begin{equation*}
	\langle \bullet, s^A_1 \rangle \smallstep{e}{A} \langle \env{\snd}{A}{B}{d}, s^A_2 \rangle
\end{equation*}
locally. This deduction step yields $p^{\sigma}_{1, 1}(n_1, n_2) = p^{\sigma}_{2, 1}(n_1, n_2)$, as the probability is not changed from state $\langle s^A_1, s^B_1 \rangle$ to $\langle s^A_2, s^B_1 \rangle$. From now on we omit the left hand side of the relation $\bigstep{}$, since it is equivalent to the right hand side of the previous deduction. We further omit writing the local deductions.

At $\langle s^A_2, s^B_1 \rangle$, globally only the [pr-e] rule can be applied. Locally, either [sys-c] or [b-c] can be applied, depending on the value of the retransmission counter $\nu_1$. Applying [sys-c] corresponds to taking the transition labelled by $\fail_1$. Since then no final state can ever be reached, we only apply the [b-c] rule locally. So we apply locally the [b-c] rule:
\begin{equation*}
	\bigstep{} \langle (\env{\snd}{A}{B}{d} + !!a_{A \rightarrow B}(d))^1, \langle s^A_3, s^B_1 \rangle, A \rangle.
\end{equation*}
This transition leads to $p^{\sigma}_{2, 1}(n_1, n_2) = p^{\sigma}_{3, 1}(n_1, n_2)$. At this point the transmission medium is invoked and globally both the [trans] and [drop] rules can be applied. If [trans] is applied, $M_B$ receives the message and we apply the [r-sys] rule locally. If [drop] is applied, merely the probability and prioritization changes. Hence, we can deduce either
\begin{align*}
\bigstep{} &\langle (\rho + ?a_{B \leftarrow A}(d) + \sys{\snd}{B}{A}{d})^{(1 - \d)}, \langle s^A_3, s^B_2 \rangle, B \rangle,~\text{or}\\
\bigstep{} &\langle (\rho)^{\d}, \langle s^A_3, s^B_1 \rangle, B \rangle,
\end{align*}
where $\rho = \env{\snd}{A}{B}{d}$. Since two transitions may be taken, we get $p^{\sigma}_{3, 1}(n_1, n_2) = (1-\d)p^{\sigma}_{3, 2}(n_1, n_2) + \d\overline{p}^{\sigma}_{3, 1}(n_1, n_2)$. After the application of [drop], $M_B$ is prioritized but cannot make a transition. In state $\langle s^A_3, s^B_1 \rangle$, globally only [npr] can be applied, with $M_A$ making a timeout transition using [to-upd] locally:
\begin{equation*}
	\bigstep{} \langle (\env{\snd}{A}{B}{d} + \timeout_1)^{\d}, \langle s^A_2, s^B_1 \rangle, A \rangle.
\end{equation*}
Since the application of [to-upd] increases the retransmission counter $\nu_1$ by one, we get $\overline{p}^{\sigma}_{3, 1}(n_1, n_2) = p^{\sigma}_{2, 1}(n_1 - 1, n_2)$ and the base case $\overline{p}^{\sigma}_{3, 1}(0, n_2) = 0$. This indicates that in state $\langle s^A_2, s^B_1 \rangle$, the deduction may be repeated with the bound $n_1$ decreased by one, corresponding to a retransmission. If $n_1 = 0$, i.e.\ in the base case, no more retransmissions are possible.

In state $\langle s^A_3, s^B_2 \rangle$ after the transmission, $M_B$ makes a transition in response to a call from its ASC. By applying [env] instead of [env$'$], we model that a is made immediately. Applying [pr-e] globally and [env] locally yields:
\begin{equation*}
	\bigstep{} \langle (\rho + \eenv{\ack}{B}{A})^{(1 - \d)}, \langle s^A_3, s^B_3 \rangle, B \rangle,
\end{equation*}
where $\rho = \env{\snd}{A}{B}{d} + ?a_{B \leftarrow A}(d) + \sys{\snd}{B}{A}{d}$. Then only [pr-e] with [b-c] can be applied (because again, applying [sys-c] does not conform with wanting to reach a final state). Therefore we get
\begin{equation}
	\bigstep{} \langle (\rho + !!b_{B \rightarrow A})^{(1 - \d)}, \langle s^A_3, s^B_5 \rangle, B \rangle,
	\label{eq:35}
\end{equation}
where $\rho = \env{\snd}{A}{B}{d} + ?a_{B \leftarrow A}(d) + \sys{\snd}{B}{A}{d} + \eenv{\ack}{B}{A}$. This generates the equalities $p^{\sigma}_{3, 2}(n_1, n_2) = p^{\sigma}_{3, 3}(n_1, n_2)$ and $p^{\sigma}_{3, 3}(n_1, n_2) = p^{\sigma}_{3, 5}(n_1, n_2)$.

In $\langle s^A_3, s^B_5 \rangle$ we can apply either [trans] globally with [r-sys] locally on $M_A$, modelling a successful transmission, or we apply [drop] globally, modelling a dropped message. Hence we can either deduce
\begin{align*}
	\bigstep{} &\langle (\rho + ?b_{A \leftarrow B})^{(1 - \d)(1 - \d)}, \langle s^A_5, s^B_5 \rangle, A \rangle,~\text{or}\\
	\bigstep{} &\langle (\rho)^{\d(1 - \d)}, \langle s^A_3, s^B_5 \rangle, A \rangle
\end{align*}
where $\rho = \env{\snd}{A}{B}{d} + ?a_{B \leftarrow A}(d) + \sys{\snd}{B}{A}{d} + \eenv{\ack}{B}{A}$. We get $p^{\sigma}_{3, 5}(n_1, n_2) = (1-\d)p^{\sigma}_{5, 5}(n_1, n_2) + \d \overline{p}^{\sigma}_{3, 5}(n_1, n_2)$. In state $\langle s^A_5, s^B_5 \rangle$, the sequence has been synchronized successfully. Here only [npr] with [to-sys] on $M_B$ can be applied to yield
\begin{equation*}
	\bigstep{} \langle (\rho + \timeout_2 + \success_2)^{(1 - \d)(1 - \d)}, \langle s^A_5, s^B_5 \rangle, A \rangle,\\
\end{equation*}
where $\rho =\env{\snd}{A}{B}{d} + ?a_{B \leftarrow A}(d) + \sys{\snd}{B}{A}{d} + \eenv{\ack}{B}{A} + $ $?b_{A \leftarrow B}$. This deduction ends in a globally final state and hence $p^{\sigma}_{5, 5}(n_1, n_2) = 1$, because the sequence $\sigma$ is correctly synchronized. In state $\langle s^A_3, s^B_5 \rangle$ after the message has been dropped, only [pr-t] with [to-upd] locally on $M_A$ can be applied:
\begin{equation*}
	\bigstep{} \langle (\rho + \timeout_1)^{\d(1 - \d)}, \langle s^A_2, s^B_5 \rangle, A \rangle
\end{equation*}
where $\rho = \env{\snd}{A}{B}{d} + ?a_{B \leftarrow A}(d) + \sys{\snd}{B}{A}{d} + \eenv{\ack}{B}{A}$. This step yields $\overline{p}^{\sigma}_{3, 5}(n_1, n_2) = p^{\sigma}_{2, 5}(n_1 - 1, n_2)$ with base case $\overline{p}^{\sigma}_{3, 5}(0, n_2) = 0$. Now $M_A$ retransmits (if its retransmission count is not yet exceeded) and we deduce with [pr-e] and [b-c] locally on $M_A$:
\begin{equation*}
	\bigstep{} \langle (\rho + !!a_{A \rightarrow B}(d))^{\d(1 - \d)}, \langle s^A_3, s^B_5 \rangle, A \rangle
\end{equation*}
where $\rho = \env{\snd}{A}{B}{d} + ?a_{B \leftarrow A}(d) + \sys{\snd}{B}{A}{d} + \eenv{\ack}{B}{A} + \timeout_1$. This yields $p^{\sigma}_{2, 5}(n_1, n_2) = \overline{\overline{p}}^{\sigma}_{3, 5}(n_1, n_2)$. Now [trans] can be applied with [r-upd] locally on $M_B$. However, when applying [drop], no final state can be reached by any sequence of applications of deduction rules. Hence we only apply [trans] and [r-upd] and get
\begin{align*}
	\bigstep{} \langle (\rho + ?a_{A \rightarrow B}(d))^{\d(1 - \d)(1 - \d)}, \langle s^A_3, s^B_5 \rangle, A \rangle
\end{align*}
where $\rho = \env{\snd}{A}{B}{d} + ?a_{B \leftarrow A}(d) + \sys{\snd}{B}{A}{d} + \eenv{\ack}{B}{A} + \timeout_1$. This yields $\overline{\overline{p}}^{\sigma}_{3, 5}(n_1, n_2) = (1 - \d)p^{\sigma}_{3, 3}(n_1, n_2 - 1)$ with base case $\overline{\overline{p}}^{\sigma}_{3, 5}(n_1, 0) = 0$.

\subsubsection{Optimization Problem}

For notational convenience, we drop the $\sigma$ superscript if the context is clear and we are not referring to a particular sequence. When the sequence of global events $\sigma$ has exactly two elements ($\len{\sigma} = 2$), we get
\newcommand{\dd}{\varrho}
\begin{align*}
	P(n_1, n_2) &\triangleq \dd(1-\d^{n_1+1}) + \frac{\dd^3}{1-\d\dd}\sum_{i=1}^{n_1}{\d^i\left[1-(\d\dd)^{M}\right]},
\end{align*}
where $\dd = (1-\d)$ is the {\define reception probability} and $M = \min{(n_1+1-i, n_2)}$. When $\sigma$ has more than two elements ($\len{\sigma} > 2$), the synchronization probability can be similarly deduced:
\begin{align*}
	P(n_1, n_2, \ldots, n_l) &\triangleq \dd \sum_{i = 0}^{n_1}{\d^{i}\locev{P}(n_1 - i, n_2 \ldots, n_l)}, \\
	\locev{P}(n_1, n_2, \ldots, n_l) &\triangleq \begin{cases}\dd \sum_{i = 0}^{M}{\d^{i}\locev{P}(n_2 - i, \ldots, n_l)} &\text{if}~l > 2 \\
		P(n_1, n_2) &\text{if}~l = 2, \\	
	\end{cases}
\end{align*}
where $M = \min{(n_1, n_2)}$. Ideally, we want to find the smallest retransmission bounds that ensure correctness. Each $p$-sequence $(\sigma)^p$ that satisfies the specification $\v$ induces a condition on the retransmission bounds associated with the elements of $\sigma$. For example, a sequence $(\e_1\e_2\ldots\e_l)^p$ induces the condition  $P^{\e_1\e_2\ldots\e_l}(n_{\e_1}, n_{\e_2}, \ldots, n_{\e_l}) \geq p$. This inequality ensures that the sequence $\sigma$ is generated by the CSAs with high enough probability as required by the correctness criterion set out above.

We can find the retransmission bounds by solving an optimization problem:
\begin{align*}
	\text{(OPT)}\hspace{0.5cm}&\min_{n_{\e_1}, n_{\e_2}, \ldots, n_{\e_l}} \sum_{j = 1}^{l} n_{\e_j}\\
	&\hspace{0.5cm}\text{s.t.}~P^\sigma(n_\sigma) \geq p \hspace{0.5cm}\text{for all}~ (\sigma)^p \in \mathcal{S_{\v}},
\end{align*}
where $\mathcal{S_{\v}} = \{ (\sigma)^p | (\sigma)^p \models \v\}$ is the set of $p$-sequences $\sigma$ that satisfy the protocol specification $\v$.

\begin{excont}{Continued}
Checking realizability of a specification amounts to checking well-posedness of the specification and feasibility of the optimization problem. For our example specification \eqref{eq:protspec}, the optimization problem is
\begin{align*}
	\min_{n_{\snd}, n_{\ack}, n_{\nack}} &n_{\snd} + n_{\ack} + n_{\nack}\\
	&\text{s.t.}~P(n_{\snd}, n_{\ack}) \geq p_1 \\
	&\phantom{\text{s.t.}}~P(n_{\snd}, n_{\nack}) \geq p_2.
\end{align*}
In the case that $p_1 = 0.7$, $p_2 = 0.8$ and $\d = 0.35$, we get $n_{\snd} = 3$, $n_{\ack} = 1$, and $n_{\nack} = 2$.
\end{excont}

\subsection{Correctness of Synthesis}

Take any protocol specification $\v$, drop probability bound $\d$, and any $p$-sequence $\sigma$ for which $(\sigma)^p \models \v$ holds. Then, correctness of the synthesis method is established by showing that $(\sigma)^{r(\sigma, \d, \M)} \models \v$, where $\M$ is the result of synthesis.

The definition of the feasible region of the optimization problem (OPT) contains the inequality $P(n_{\sigma}) \geq p$ for each such sequence $\sigma$. By the semantics of protocol specifications $(q \geq p \wedge (\sigma)^{p} \models \v) \Rightarrow (\sigma)^{q} \models \v$ for any sequence $\sigma$. It is therefore sufficient to show that $(\sigma)^{P(n_{\sigma})} \models \v$ and $r(\sigma, \d, \M) \geq P(n_\sigma)$, because then $(\sigma)^{r(\sigma, \d, \M)} \models \v$, as required to establish correctness.

First, if the retransmission bounds $n_{\sigma}$ are part of a feasible solution to (OPT), then we necessarily have $P(n_{\sigma}) \geq p$, and so $(\sigma)^{P(n_{\sigma})} \models \v$ follows from $(q \geq p \wedge (\sigma)^{p} \models \v) \Rightarrow (\sigma)^{q} \models \v$.

Second, we have $P(n_\sigma) = r(\sigma, \d, \M)$ by construction of $P$ (note that the superscript $\sigma$ has been dropped from $P^{\sigma}$): $r(\sigma, \d, \M)$ is the sum of all probabilities $p$ for which $\proj{\rho} = \sigma \wedge (\rho)^p \generated$, i.e.\ the environment-triggered events and system-triggered events in $\rho$ synchronize to the sequence of global events $\sigma$ and the $p$-sequence $(\rho)^p$ is generated by $\M$ and drop probability $\d$. By definition, $(\rho)^p \generated \Leftrightarrow \exists s^f \in S^f_{\M} . \exists x, y \in \SAP . \langle (\bullet)^1, s^{init}, x \rangle \bigstep{}^* \langle (\rho)^p, s^f, y \rangle$. It is therefore sufficient to show that in the deduction of the expression for $P(n_{\sigma})$ exactly those $p$-sequences $(\rho)^p$ are taken into account that end in a globally final state $s^f \in S^f_{\M}$ (the prioritization of $x$ and $y$ can safely be ignored) and for which $\proj{\rho} = \sigma$.

The deduction of $P(n_{\sigma})$ in \sec{Pdeduction} is essentially done by constructing a product automaton of all CSAs in $\M$ using the global semantics, and adding the probabilities along all paths that end in a globally final state corresponding to the sequence $\sigma$ having been executed.

Note that it would have been enough to show $P(n_\sigma) \leq r(\sigma, \d, \M)$. A synthesis method that generates CSAs with $P(n_\sigma) = 0$ would be perfectly correct, but not very useful: The larger $P(n_\sigma)$ gets, the greater the feasible region of (OPT) gets and the more specifications can be synthesized. So by having $P(n_\sigma) = r(\sigma, \d, \M)$, we have maximized the capabilities of the synthesis method.

\subsection{Computational Considerations} \label{sec:computation}

The time required to generate a CSA from a protocol specification $\v$ by the {\sc{Synthesize}} algorithm is proportional to the number of global events and disjunctions ($\vee$) in $\v$ (ignoring the set operations on $E$), which can easily be seen from \tab{synthesis}, where the implementation of {\sc{Synthesize}} is shown as a simple structural recursion on $\v$. When also the set operations on $E$ are taken into account, the algorithm is quadratic in the number of global events in $\v$.

The main computational complexity arises from the optimization problem OPT, which is an integer program and in general is NP-hard. There are however a few points to be noted that may simplify finding a solution. First, both the objective function and the function $P^{\sigma}(n_{\sigma})$ are monotonous in their arguments along any dimension. Hence, if OPT is feasible for some $n = (n_{\e_1}, n_{\e_2}, \ldots, n_{\e_l})$, it is also feasible for any $n' \geq n$.

Second, since correctness depends on $P^{\sigma}(n_{\sigma}) \leq r(\sigma, \delta, \M)$, it is sufficient to solve an optimization problem with a strictly smaller feasible set than that of (OPT). This is helpful if a function $Q^{\sigma}$ can be found s.t.\ for all $\sigma$, $Q^{\sigma}(n_{\sigma}) \leq P^{\sigma}(n_{\sigma})$ while still maintaining that there exist retransmission bounds $n_{\sigma}$ s.t.\ $Q^{\sigma}(n_{\sigma}) \geq p$ for all $(\sigma)^p \in S_\v$. The solution to the resulting optimization problem might not be optimal, but the resulting CSAs are still correct.

Lastly, since any suboptimal solution to OPT still gives rise to correct CSAs,  the retransmission bounds may be chosen to be arbitrarily high as long as they are feasible. Note however that there might not be a solution to OPT at all, in which case the specification was unrealizable in the first place.

\subsection{Discussion} \label{sec:validity}

\begin{figure}
\centering
	\psfrag{data length}[cc][cc]{data length $d_{max}$}
	\psfrag{number of cars}[rb][rt]{$\begin{array}{l}\text{number}\\\text{of cars $N$}\end{array}$}
	\psfrag{minimum delay}[cc][cc]{minimum delay $\tau_{min}$}
	\includegraphics[width=0.45\textwidth]{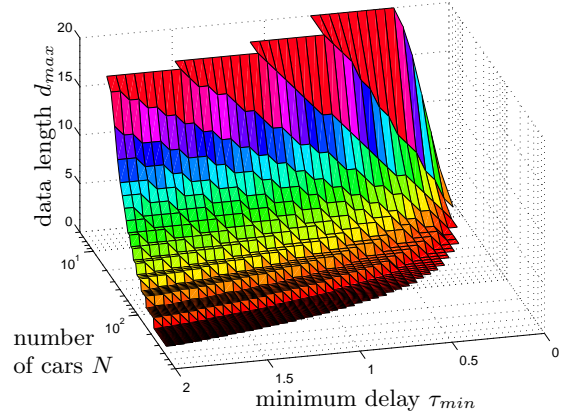}
\figsmallvspace
\caption{Feasible region of OPT for the protocol specification in \eqref{eq:protspec} with $p_1 = p_2 = 0.9$. The drop probability bound $\d$ is calculated as a function of $N$, $d_{max}$ and $\tau_{min}$. All points on and under the surface are feasible.}
\label{fig:feasibility6}
\figvspace
\end{figure}

The implementations of a communication protocol specification provide the ASCs with sufficient information on what messages are received so that accidents can effectively be prevented. In this section we develop the continuing example of the cars at an intersection further by explaining how our protocol can be embedded in an active safety application.

\begin{excont}{Continued}
When transmitting data $d$ from car $A$ to car $B$, six cases can occur. We distinguish the cases by the final system-triggered events that generate upcalls to the ASCs on either car. The case we call ``correct'' is when $B$ receives $d$, $A$ knows about it and $B$ assumes correctly that $A$ knows. $B$ then correctly receives a ``$\success$'' upcall, which is consistent with $A$'s last upcall. The ASCs can then correctly react in a consistent way, e.g.\ by one car gracefully decelerating.

In other cases the ASCs can still react in a safe way even if $A$ and $B$ have inconsistent information about each other: If $B$ receives $d$ correctly, $A$ never receives an acknowledgement and $B$ assumes $A$ never did, then both ASCs receive ``$\fail$'' upcalls and can react accordingly. If $B$ receives $d$ correctly and $A$ receives the acknowledgement, but $B$ assumes $A$ did not receive it, then $B$ receives a ``$\fail$'' upcall and can react conservatively. If $B$ does not receive $d$ and $A$ holds that it did not, then the ASC on $A$ can react conservative on its ``$\fail$'' upcall. If $B$ receives $d$ correctly, $A$ misses the acknowledgement but $B$ holds that $A$ received it, then $A$ incorrectly assumes the worst case but yet reacts conservatively.

The only problematic case is when $B$ does not receive $d$ but $A$ holds that it did. Then the ASC on neither $A$ nor $B$  takes conservative action, potentially resulting in an accident. However, the synthesis method constructs the CSAs so that this case never occurs under the given assumptions.
\end{excont}

We now conclude the example by presenting numerical results that illustrate in which hypothetical scenarios protocols that we are considering are realizable.

\begin{excont}{Continued}
As introduced above, the drop probability bound $\d$ on the transmission medium may be calculated from other more readily available parameters. The realizability of a given protocol specification $\v$ depends on the drop probability bound $\d$. For demonstrative purposes, we calculate $\d$ from the number of cars $N$ at the intersection that may use the transmission medium simultaneously, the minimum time $\tau_{min}$ it may take for a message to be sent between two cars and the maximum amount of data $d_{max}$ that may be carried in a message. Given an empirically obtained function $\d(r)$ that maps a data-rate $r$ to a drop-probability of the transmission medium, we calculate $\d(r)$ with $r = (N-2)d_{max}/\tau_{min}$ (we take $N-2$ as we consider the environment to be all cars except the two that are communicating).

We illustrate the effectiveness of our synthesis method by asserting the sigmoid $\d(r) = (1 + a\cdot\exp(-br))$ with $a = 4$ and $b = 0.002$. Using the protocol specification in \eqref{eq:protspec}, we illustrate how realizability changes with different values for the number of cars $N$, minimum time to deliver a message $\tau_{min}$ and maximum amount of data in a message $d_{max}$. \fig{feasibility6} shows the feasible region of (OPT) for $\v$ with $p_1 = p_2 = 0.9$, i.e.\ for which values of $\d$ calculated as a function of $N$, $\tau_{min}$ and $d_{max}$ the synthesis problem is realizable.

It is clearly visible from \fig{feasibility6} that the more cars are sharing the transmission medium, the smaller the delay, and the larger the packets, the higher the worst-case data rate could be on the network, and the specification becomes harder to realize. If moreover the requirements $p_1$ and $p_2$ are made more stringent, the feasible region decreases even further.
\end{excont}

\vspace{-.06in}

\section{Conclusion} \label{sec:conclusion}

This work demonstrates a framework for reliable communication protocols for intervehicular communication in active safety applications. The framework, consisting of a precisely defined specification language and execution model (in the form of CSAs), allows for correct-by-construction synthesis of protocol implementations that satisfy the specifications even in the presence of several other cars sharing the transmission medium.

In our synthesis method we only take into account the drop probability of the transmission medium and assume that this is sufficient to synthesise reliable protocols. This also only enables to guarantee QoS requirements on the reception probability. Furthermore, in the current formulation, only two cars can participate in a dialogue, but some active safety applications might require to extend this. Also, note that if a communication is under way, the arrival of another message cannot directly be handled even if it is required to satisfy the QoS requirements.

Our approach permits several extensions: (i) Allowing the higher level to specify the QoS requirements and the destination address at runtime (i.e.\ for each transmission), (ii) Guaranteeing QoS requirements on the end-to-end delay of the communication and more general assumptions on the transmission medium dynamics in order to widen the range of applicability, and (iii) including the capability to relay messages over several cars to create a routed network. The latter would also require a rigorously developed synthesis method for protocols to discover the network topology, which we are currently working on.\\

\vspace{-.06in}

\section*{Acknowledgements} The authors would like to extend thanks to Rohit Pandita and Vladimeros Vladimerou from Toyota as well as Scott Livingston, Pavithra Prabhakar and Eric Wolff at the California Institute of Technology for fruitful discussions.

\bibliographystyle{abbrv}
\small{
\bibliography{iccps2013-wiltsche}
}

\end{document}